\documentclass[]{aastex631}

\shorttitle{Mass loss from extreme planets}
\shortauthors{Koskinen et al.}
\graphicspath{{./}{figures/}}
\begin{document}

\title{Mass loss by atmospheric escape from extremely close-in planets}%\footnote{Released on March, 1st, 2021}}

\author{Tommi T. Koskinen}
\affiliation{Lunar and Planetary Laboratory, University of Arizona \\
1629 E. University Blvd. \\
Tucson, AZ 85721--0092, USA}
\affiliation{Alien Earths Team, NASA Nexus for Exoplanet System Science, University of Arizona, Tucson, AZ, USA}

\author{Panayotis Lavvas}
\affiliation{Groupe de Spectrom\'etrie Mol\'eculaire et Atmosph\'erique UMR CNRS 6089 \\
Universit\'e Reims Champagne-Ardenne \\
51687 Reims, France}

\author{Chenliang Huang}
\affiliation{Lunar and Planetary Laboratory, University of Arizona \\
1629 E. University Blvd. \\
Tucson, AZ 85721--0092, USA}

\author{Galen Bergsten}
\affiliation{Lunar and Planetary Laboratory, University of Arizona \\
1629 E. University Blvd. \\
Tucson, AZ 85721--0092, USA}
\affiliation{Alien Earths Team, NASA Nexus for Exoplanet System Science, University of Arizona, Tucson, AZ, USA}

\author{Rachel B. Fernandes}
\affiliation{Lunar and Planetary Laboratory, University of Arizona \\
1629 E. University Blvd. \\
Tucson, AZ 85721--0092, USA}
\affiliation{Alien Earths Team, NASA Nexus for Exoplanet System Science, University of Arizona, Tucson, AZ, USA}

\author{Mitchell E. Young}
\affiliation{Department of Physics, University of Oxford \\
Denys Wilkinson Building, Keble Road \\
Oxford, OX1 3RH, UK}



\begin{abstract}
\noindent
We explore atmospheric escape from close-in exoplanets with the highest mass loss rates. First, we locate the transition from stellar X-ray and UV-driven escape to rapid Roche lobe overflow, which occurs once the 10--100 nbar pressure level in the atmosphere reaches the Roche lobe. Planets enter this regime when the ratio of the substellar radius to the polar radius along the visible surface pressure level, that aligns with a surface of constant Roche potential, is X/Z~$\gtrsim$~1.2 for Jovian planets (Mp~$\gtrsim$~100 M$_{\Earth}$) and X/Z~$\gtrsim$~1.02 for sub-Jovian planets ($M_p \approx$~10--100 M$_{\Earth}$). Around a sun-like star, this regime applies to orbital periods of less than two days for planets with radii of about 3--14 R$_{\Earth}$. Our results agree with the properties of known transiting planets and can explain parts of the sub-Jovian desert in the population of known exoplanets. Second, we present detailed numerical simulations of atmospheric escape from a planet like Uranus or Neptune orbiting close to a sun-like star that support the results above and point to interesting qualitative differences between hot Jupiters and sub-Jovian planets. We find that hot Neptunes with solar metallicity hydrogen and helium envelopes have relatively more extended upper atmospheres than typical hot Jupiters, with a lower ionization fraction and higher abundances of escaping molecules. This is consistent with existing ultraviolet transit observations of warm Neptunes and it might provide a way to use future observations and models to distinguish solar metallicity atmospheres from higher metallicity atmospheres. 
\end{abstract}



\section{Introduction} \label{sc:intro}

\noindent
Atmospheric escape can have significant consequences for the evolution of planetary atmospheres as well as the structure and bulk density of different planets. For example, it is widely accepted that the early atmospheres of Venus, Earth, and Mars were shaped by escape \citep[e.g.,][]{lammer08}. Exoplanets offer an extended parameter space that we can use to test models of atmospheric escape. The population of known exoplanets includes features such as the radius valley between rocky planets and planets with a more substantial gaseous envelope \citep{fulton17}, and the hot Neptune/Super-Earth deserts \citep[e.g.,][]{mazeh16,berger18} that are often linked to the loss of gaseous envelopes over time \citep[e.g.,][]{owen17,gupta20}. Many transiting exoplanets also exhibit signatures of atmospheric escape in their transmission spectra \citep[e.g.,][]{vidalmadjar03,linsky10,lecavelier12,benjaffel13,ehrenreich15,bourrier18,spake18,sing19} that have inspired new developments in the theory of atmospheric escape. 

The implications of atmospheric escape on the envelopes of close-in planets have been debated since the first hot Jupiter, 51 Pegasi b, was discovered \citep{mayor95}. Since then, observations of transiting planet atmospheres have constrained the escape mechanism and mass loss rates in theoretical models and revealed qualitative differences between different types of planets and host stars. For example, transit observations in the far-ultraviolet (FUV) obtained by the Hubble Space Telescope (HST) probe hydrogen in the escaping upper atmosphere, along with some heavier elements such as carbon, oxygen, and silicon if the latter are present at high altitudes with sufficient abundances. These observations show that many close-in planets lose mass to hydrodynamic escape. They also indicate that Neptune-mass planets such as GJ436b and GJ3470b have larger upper atmosphere transit depths \citep{ehrenreich15,bourrier18} than hot Jupiters such as HD209458b or HD189733b \citep{vidalmadjar03,lecavelier12}.   

Generally, escape is powered by ionization and heating of the planetary upper atmosphere by stellar X-ray and UV (XUV) radiation. Most models of this process agree that typical hot Jupiters do not lose a substantial fraction of their envelopes to mass loss, even when their atmospheres escape hydrodynamically. Detailed models of the upper atmosphere that have been used to explain transit observations generally predict mass loss rates of the order of 10$^7$--10$^8$ kg~s$^{-1}$ for `mainstream' hot Jupiters such as HD209458b or HD189733b \citep[e.g.,][]{yelle04,yelle06,garciamunoz07,koskinen13a,koskinen13b,chadney17}. With these rates, cumulative mass loss from many hot Jupiters amounts to at most a few percent of the current planet mass, depending on the XUV evolution of the host star. This conclusion, however, does not necessarily apply to all hot Jupiters. 

Extremely close-in planets such as WASP-12b and WASP-121b have near-UV (NUV) transit depths that substantially exceed the visual transit depth and therefore probe the escaping upper atmosphere \citep{fossati10,sing19}. Since hydrogen and helium do not absorb in the NUV, the transit depths are due to heavier elements and metals that escape the atmosphere in large quantities. As might be expected for such planets, this implies exceptionally high mass loss rates. These planets orbit so close to their host stars, with an orbital period of just $\sim$1 day, that their envelopes are substantially distorted by the stellar tide. Therefore, they fill a significant fraction of their Roche lobe i.e., the smallest equipotential surface that can exist around the planet in the combined gravity field of the planet and the star. 
Under these circumstances, the middle atmosphere extends to the Roche lobe and the resulting high mass loss rates can have cataclysmic consequences \citep{li10,lai10}. In this case, mass loss is powered by the stellar bolometric luminosity and depends roughly on the effective temperature of the planet instead of the stellar XUV radiation that heats the upper atmosphere. 

Observations of transiting warm Neptunes (10--25 $M_{\Earth}$), at Lyman~$\alpha$ and the He I 1083 nm line, imply mass loss rates of about 10$^6$--10$^9$ kg~s$^{-1}$ i.e., of the same order of magnitude as for the known hot Jupiters \citep[e.g.,][]{ehrenreich15,parkeloyd17,bourrier18,mansfield18}. While mass loss rates ostensibly retrieved from observations are actually model-dependent in the sense that they depend on what physics is included in the model, and on the assumed planet and host star properties, these values are consistent with the idea that the atmospheres of hot Neptunes and lower-mass planets in general are even more likely to undergo hydrodynamic escape than the atmospheres of hot Jupiters \citep{koskinen14}. If this is the case, the mass loss rate is `energy-limited' \citep[e.g.,][]{watson81,erkaev07} roughly to within an order of magnitude, and the escape rate depends inversely on the bulk density of the planet, which is similar for hot Neptunes and Jupiters. The relative impact of mass loss on hot Neptunes, however, is larger, especially if one accounts for their evolution over time and the higher stellar XUV fluxes of young stars \citep[e.g.,][]{lopez12}. Hot Neptunes and lower-mass planets are also more susceptible to mass loss by Roche lobe overflow than hot Jupiters. Lower-mass planets may therefore lose a substantial fraction of their gaseous envelope over time, explaining the occurrence of the hot Neptune/Super-Earth desert in the population of known planets \citep[e.g.,][]{owen17}.   

In this work, we focus on close-in planets that are likely to have the highest mass loss rates. In particular, we explore the transition from stellar XUV-driven escape of the upper atmosphere to rapid Roche lobe overflow of the lower and middle atmosphere. This transition relates to a fundamental stability limit for gaseous envelopes. We present a relatively simple theory for it, highlight the underlying physics, and show how the equilibrium shape of a planet can be used to assess the stability of its envelope. While Roche lobe overflow is a well known phenomenon in binary star, exoplanet, and planetary systems in general, the results here point to some new insights to the nature of its application to close-in exoplanets. In binary star systems, for example, Roche lobe overflow occurs when one of the stars fills its Roche lobe and begins to lose mass to the other star. The same is expected to occur in close-in exoplanet systems where the material lost from the planet forms a torus around the star \citep{fossati13}, analogous to, say, the Io plasma torus around Jupiter. Near the classical Roche limit, the surface of the planet reaches the Roche lobe and the planet loses its envelope to cataclysmic mass loss. The visible surface need not fill the Roche lobe, however, for the planet to lose its envelope over a relatively short time \citep[e.g.,][]{jackson17}. Here, we quantify the limit for significant envelope loss in terms of the tidal perturbation of the planet and extent of its atmosphere, and properly relate it to stellar XUV-driven escape. 

We compare our predictions with detailed simulations atmospheric escape and validate them by comparison with the observed properties of known exoplanets. Since Neptune-mass planets undergo rapid mass loss more readily than hot Jupiters, we focus the detailed simulations on a Uranus-like planet orbiting close-in to a sun-like star. An additional purpose of these simulations is to identify qualitative differences between planets similar to Uranus and Neptune and previously published simulations of hot Jupiters. This helps to inform approximations of mass loss used in other studies and consistently explain some of the general differences between transit observations of these types of planets. Here, we model a Uranus-like planet because its mass, radius, and overall structure are consistent with and represent a real planet, instead of depending on uncertain models of interior structure and composition. We chose Uranus instead of Neptune because its surface gravitational potential is slightly lower than that of Neptune, resulting in a somewhat more extended atmosphere that might be more typical of close-in planets.  

\section{Methods}
\label{sc:methods}

\noindent
Below, we provide a relatively concise summary of our methods and underlying theory. In Section~\ref{sc:shape}, we explain how we include the Roche potential in our models and calculate the shape of the equipotential surfaces for different planets. In Section~\ref{subsc:lowatm}, we present a straightforward method for estimating the extent of the lower and middle atmosphere that affects the mass loss rates. In sections~\ref{subsc:overflow1} and \ref{subsc:energy_limit}, we provide a simple approximation for calculating mass loss rates due to Roche lobe overflow and review the equations for energy-limited escape that are often used to represent stellar XUV-driven escape \citep[see][for review]{owen20}. In Section~\ref{subsc:escape_model} and Appendix~\ref{ap:atmosmodel}, we describe a numerical, multi-species model of hydrodynamic escape that we use to test and validate our simple equations of atmospheric escape, in addition to comparing our predictions with the properties of known exoplanets. 

\subsection{Equilibrium shape}
\label{sc:shape}

\noindent
At close-in orbits, pressure levels in the gaseous envelopes of planets are significantly perturbed by the stellar gravity field. As the orbital distance decreases, the outer layers of the atmosphere and eventually the visible surface of the planet approach the Roche lobe. In hydrostatic equilibrium, isobars coincide with surfaces of constant gravitational potential, provided that horizontal variations in temperature can be ignored. The relevant gravitational potential here is the Roche potential \citep[e.g.,][]{pringle85}:
\begin{eqnarray}
U (x,y,z) &=& -\frac{GM_p}{(x^2+y^2+z^2)^{1/2}} - \frac{GM_s}{\left[ (x-a)^2+y^2+z^2 \right]^{1/2}} \nonumber \\
& &- \frac{1}{2} \Omega^2 \left[ (x-\mu a)^2 + y^2 \right]
\label{eqn:rochepot}
\end{eqnarray}    
where $z$ is the distance along a coordinate axis pointing to the north pole from the center of the planet, $x$ axis points towards the star, $y$ axis completes a right-handed coordinate system, $G$ is the gravitational constant, $M_p$ is the mass of the planet, $a$ is the orbital distance, $M_s$ is the mass of the star, $\Omega$ is the orbital angular rotation rate, given by Kepler's third law: 
\begin{equation}
\Omega^2 = \frac{G (M_s+M_p)}{a^3},
\label{eqn:kepler3}
\end{equation}
and
\begin{equation}
\mu = \frac{M_s}{M_s+M_p}
\end{equation}
is the mass ratio. In this study, we assume that planets orbit their host stars in a circular orbit within 0.1 AU and are tidally locked \citep[e.g.,][]{guillot96}. In spherical polar coordinates centered at the planet, the Roche potential is
\begin{eqnarray}
U (r, \theta, \phi) &=& -\frac{G M_p}{r} - \frac{G M_s}{\left[ \left( a - r \zeta \right)^2 + r^2 \xi^2 + r^2 \nu^2 \right]} \nonumber \\
&-& \frac{1}{2} \Omega^2 \left[ \left( r \zeta - \mu a \right)^2 + r^2 \xi^2 \right] 
\end{eqnarray}
with 
\begin{eqnarray}
\zeta &=& \sin \theta \cos \phi \\
\xi &=& \sin \theta \sin \phi \\
\nu &=& \cos \theta
\end{eqnarray}
where $\theta$ and $\phi$ are co-latitude and longitude, respectively. We solve for equipotential surfaces numerically by using an iterative method (see Appendix~\ref{ap:equipotentials}). 

As we will demonstrate, the equilibrium shape of the pressure levels based on the Roche potential can be used to predict the stability of gaseous envelopes.  In Section~\ref{sc:results}, we describe the equilibrium shape by calculating the ratio of the substellar radius to the polar radius ($x/z$) for different planets. This ratio is obtained by equating
\begin{equation}
U(x,0,0) = U(0,0,z)
\label{eqn:shape1}
\end{equation}
and solving for $x/z$. The L1 point $r_{L1} = x_R$ is the substellar radius of the Roche lobe equipotential surface, for which $x_R/z_R \sim$~1.5. We note that substellar gravity is given by
\begin{equation}
g_s (r) = -\frac{\partial U(r, \pi/2, 0)}{\partial r} = -\frac{G M_p}{r^2} + \frac{G M_s}{(a - r)^2} + \Omega^2 (r - \mu a).
\label{eqn:ssgrav}
\end{equation}
We obtain the radius at the L1 point by setting the substellar gravity to zero and solving the resulting equation for $r_{L1}$ by using Newton's method. 

\subsection{Lower and middle atmosphere}
\label{subsc:lowatm}    

\noindent
In addition to the gravity field, the extent of the lower and middle atmosphere depends on temperature and composition through mean molecular weight. State of the art models for the composition and thermal structure of the lower and middle atmosphere include self-consistent treatments of photochemistry/thermochemistry and radiative transfer in the presence of clouds and hazes \citep[e.g.,][]{lavvas21}. We do not aim to reproduce such complexity here. Instead, we develop a simple approximation to estimate the extent of the lower and middle atmosphere for a H/He envelope to highlight its effect on mass loss. This approximation is roughly valid for a solar metallicity gas where the mean molecular weight is largely set by the relative abundances of H$_2$, He, and H. While higher metallicity envelopes are possible, this assumption is consistent with most current studies of mass loss and exoplanet populations \citep[e.g.,][]{owen17}.

We calculate the extent of the atmosphere, or pressure level heights, by numerically integrating the hypsometric equation:
\begin{equation}
\int_{h_0}^h \frac{m(h') g(h') \text{d} h'}{k T(h')} = -\ln \left[ \frac{p(h)}{p(h_0)} \right] 
\label{eqn:preslevels}
\end{equation}
where $h$ is geopotential height, $dU = g(h) \text{d} h$, $h_0$ is a reference height at the lower boundary, $g(h)$ is gravity (perpendicular to equipotential surfaces), $m(h)$ is the mean molecular weight, $k$ is the Boltzmann constant, and $T(h)$ is temperature. We assume that the temperature is constant throughout the lower and middle atmosphere and equal to the effective temperature of the planet. In order to calculate the effective temperature, we use a generic Bond albedo $A_B =$~0.3 and assume uniform redistribution of energy around the planet.

\begin{figure}
  \epsscale{1.1}
  \plotone{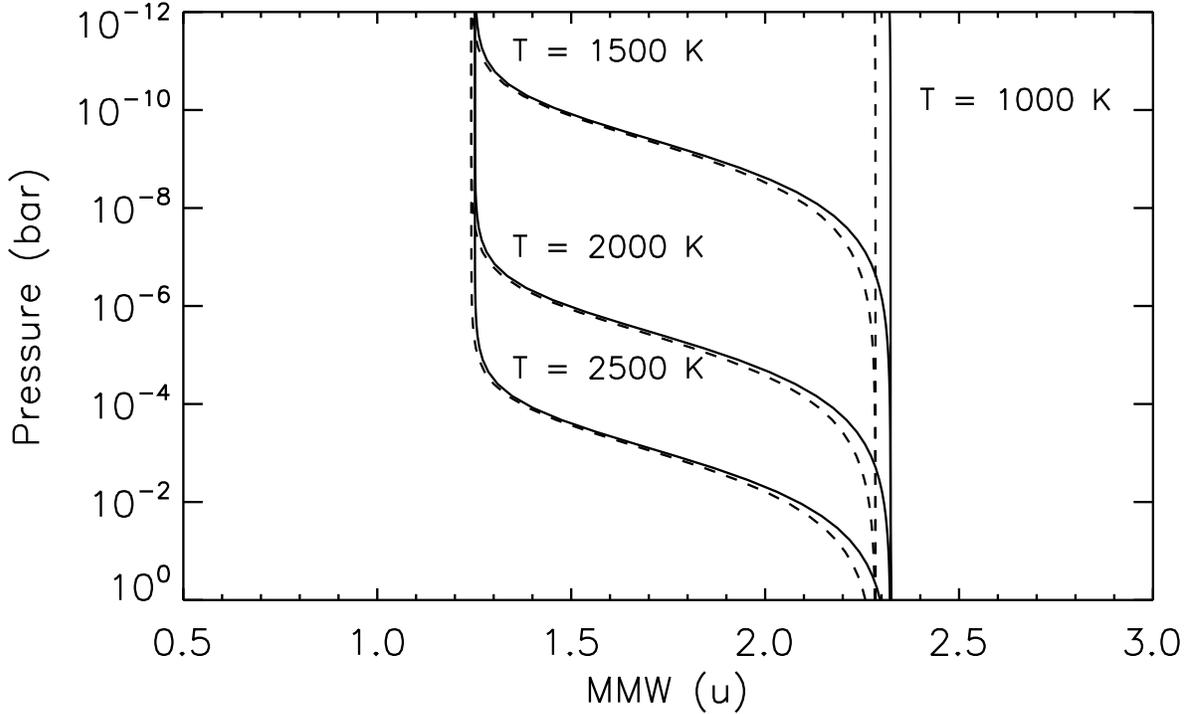}
  \caption{Mean molecular weight based on the CEA chemical equilibrium model \citep[solid lines,][]{gordon94} compared with the mean molecular weight based on the simple equations (\ref{eqn:h2x})--(\ref{eqn:hx}) (dashed lines) for effective temperatures of 1000--2500 K.}
  \label{fig:eqchem}
\end{figure}    

We assume that the mean molecular weight of the atmosphere depends only on the relative abundances of H$_2$, He, and H. For a solar metallicity gas, the volume mixing ratios of these species can be approximated as \citep{visscher06}: 
\begin{eqnarray}
q_{\text{H}_2} &\approx& \frac{1.9845 + 10^u - \sqrt{10^u (3.9690+10^u)}}{2.3670} \label{eqn:h2x} \\
u &=& -\frac{23672}{T} - \log_{10} p + 6.2645 \nonumber \\
q_{\text{He}} &=& \frac{f_{\text{He}} (1 - q_{\text{H}_2}) + 2 f_{\text{He}} q_{\text{H}_2}}{f_{\text{He}}} \\
q_{\text{H}}   &=& 1 - q_{\text{H}_2} - q_{\text{He}} \label{eqn:hx} 
\end{eqnarray}
where $p$ is pressure in bar and $f_{\text{He}} =$~7.93~$\times$~10$^{-2}$ is the abundance of He relative to H \citep{lodders03}. In order to test the validity of these approximations, we use the NASA Chemical Equilibrium Applications (CEA) code \citep{gordon94} to calculate the composition at temperatures of 1,000 K, 1,500 K, 2,000 K, and 2,500 K. Figure~\ref{fig:eqchem} compares the mean molecular weight based on the model with that based on the above equations. The lack of heavy elements leads to a slightly lower mean molecular weight in our simple model but otherwise, the agreement is relatively good. We note that our approach here ignores photochemistry. Since photolysis leads to more efficient dissociation of H$_2$ and other molecules, our results underestimate the extent of the atmosphere for cooler planets with effective temperatures of $\sim$1,000--2,000 K. At higher temperatures, the composition is expected to approach thermochemical equilibrium \citep{moses11}.

\subsection{Roche lobe overflow}
\label{subsc:overflow1}

\noindent
Roche lobe overflow occurs when the atmosphere of the planet extends to the Roche lobe \citep[e.g.,][]{lecavelier04,li10}. Otherwise, the outer limit of the atmosphere is the exobase where the mean free path between collisions exceeds the pressure scale height ($p \approx$~10$^{-12}$ bar). For our purposes, Roche lobe overflow occurs when the L1 point is located below the exobase. If the temperature and mean molecular weight do not change significantly with latitude and longitude, Roche lobe is a surface of constant pressure and density, and once the L1 point is below the exobase, the entire Roche lobe is below the exobase. 

Roche lobe overflow can support substantial mass loss even in the absence of stellar XUV heating \citep[e.g.,][]{li10}. In general, we can estimate the mass loss rate based on the underlying atmospheric structure by adapting Jeans escape for `geometrical blow-off' \citep{lecavelier04}:
\begin{equation}
\dot{M} \approx 4 \pi x_c z_c \frac{\rho_c}{2 \sqrt{\pi}} \sqrt{ \frac{2 k T_c}{m_c} } \left( \Gamma_c+1 \right) \exp(-\Gamma_c)
\label{eqn:mloss1}
\end{equation}
where the subscript $c$ refers to the exobase, $\rho_c$ is mass density, $x_c$ and $z_c$ are the substellar and polar radii, respectively, of the corresponding equipotential surface, and $\Gamma_c$ is the thermal escape parameter:
\begin{equation}
\Gamma_c = \frac{m_c \Delta U}{k T_c}.
\end{equation}
Here, $\Delta U$ is the potential difference from the exobase to the Roche lobe. This potential difference approaches zero when the Roche lobe approaches the exobase. When the Roche lobe is at the exobase or below, the mass loss rate is
\begin{equation}
\dot{M}_R \approx 4 \pi x_R z_R \frac{\rho_R}{2 \sqrt{\pi}} \sqrt{ \frac{2 k T_R}{m_R} }.
\label{eqn:mloss2}
\end{equation} 
Effectively, this means that the atmosphere escapes through the Roche lobe close to the speed of sound. In Section~\ref{subsc:num_model}, we use a detailed numerical model of hydrodynamic escape to show that this approximation holds for flow through the L1 point and argue that in principle, equation~(\ref{eqn:mloss2}) provides a reasonably good estimate of the global mass loss rate due to Roche lobe overflow.

If the night side is substantially colder than the dayside, the density at the Roche lobe boundary could be significantly lower than on the dayside. In the unlikely case of no escape at all from the night side, the mass loss rate would be reduced by a factor of two. A more conservative assumption is that Roche lobe overflow is limited to a nozzle around the L1 point. \citet{lai10} estimate the radius of this nozzle as:
\begin{equation}
\Delta R_L \approx \frac{c_s}{\Omega},
\end{equation}
where $c_s$ is the sound speed, and give the mass loss rate as:
\begin{equation}
\dot{M}_L \approx \pi \rho_R c_s \left( \Delta R_L \right)^2.
\end{equation}
For the planets considered in this paper, the mass loss rates based on this approximation are lower by factors of 2--10 compared to equation~(\ref{eqn:mloss2}) once the atmosphere enters the Roche lobe overflow regime. We conclude that equation~(\ref{eqn:mloss2}) is accurate to a factor of a few even if conditions around the planet are significantly different.

\subsection{Energy-limited escape}
\label{subsc:energy_limit}

\noindent
Energy-limited escape is often assumed in the literature to calculate mass loss due to stellar XUV-driven atmospheric escape or photoevaporation\footnote{The term photoevaporation is technically misleading because evaporation was traditionally used to describe thermal Jeans escape whereas many exoplanet models intend to simulate hydrodynamic escape.}. For this reason, we also compare our results with this approach in Section~\ref{sc:results}. The standard formula for mass loss by energy-limited photoevaporation \citep{watson81} can be written as:
\begin{equation}
\dot{M}_{\text{XUV}} = \frac{\epsilon L_{\text{XUV}}}{4 \Delta U} \left( \frac{r_e}{a} \right)^2  
\label{eqn:enerlim}
\end{equation}
where $\epsilon$ is the mass loss (or heating) efficiency, $L_{\text{XUV}}$ is the XUV luminosity of the star, $r_e$ is the effective radius in the planet's atmosphere where XUV radiation is absorbed and $\Delta U$ is the potential difference that escaping particles must overcome. This formula assumes uniform redistribution of energy around the planet. Typically, photoionization of H and He are assumed to heat the upper atmosphere and $L_{\text{XUV}}$ includes radiation at energies higher than the ionization threshold of H at 13.6 eV ($\lambda \le$~91.1 nm). The apparent simplicity of the formula, particularly the linear dependency on the stellar XUV luminosity, is deceptive. Much of the related atmospheric physics is buried in the efficiency $\epsilon$ that depends on stellar luminosity and atmospheric composition \citep[e.g.,][]{murrayclay09,koskinen14}. The lack of knowledge about atmospheric structure also makes it difficult to properly define the potential difference $\Delta U$.  

The simplest form of the equation, often adapted in studies of exoplanet evolution, assumes an isolated object for which $\Delta U = GM_p/R_p$ where $R_p$ is the radius of the planet that would be probed by visual transit observations. In this case, we have \citep[e.g.,][]{owen17}:
\begin{equation}
\dot{M}_{\text{XUV}} = \frac{\epsilon \pi R_p^3 L_{\text{XUV}}}{4 \pi a^2 G M_p}. 
\label{eqn:enerlim2}
\end{equation}
Here, energy-limited escape is inversely proportional to the bulk density of the planet. This explains why it returns similar mass loss rates for a wide range of different planets, ranging from Neptune-mass planets to Jovian planets, that tend to have similar bulk densities. Given that bulk densities vary by less than an order of magnitude in general, energy-limited mass loss rates for rocky planets are also similar in magnitude to the giant planets.

Another commonly used form of energy-limited escape includes a correction factor to approximate Roche lobe overflow. This was proposed by \citet{erkaev07} who wrote the potential difference in equation~(\ref{eqn:enerlim}) as 
\begin{equation}
\Delta U \approx \frac{GM_p}{R_p} \left( 1 - \frac{3}{2 \eta} + \frac{1}{2 \eta^3} \right) = \frac{GM_p}{R_p} K(\eta)
\label{eqn:erkaevdu}
\end{equation}  
where $\eta = x_R/R_p$ and the L1 point is at
\begin{equation}
x_R \approx \left( \frac{\delta}{3} \right)^{1/3} \left[ 1 - \frac{1}{3} \left( \frac{\delta}{3} \right)^{1/3} \right] a
\label{eqn:erkaevl1}
\end{equation} 
where $\delta = M_p/M_s$. This approach assumes that both the surface of the planet at $r = R_p$ and the Roche lobe are surfaces of constant density and that the atmosphere escapes uniformly through the Roche lobe. In line with equation~(\ref{eqn:enerlim2}), it does not account for the extent of the atmosphere.
 
\subsection{Upper atmosphere escape model}
\label{subsc:escape_model}

\noindent
In order to test the simplified formulae for Roche lobe overflow and energy-limited photoevaporation, and to capture the transition from photoevaporation to Roche lobe overflow explicitly, we use a detailed, one-dimensional model of hydrodynamic escape to simulate the upper atmosphere. In addition to the brief overview below, Appendix~\ref{ap:atmosmodel} summarizes the details of the model. Our intent here is not to simulate any known exoplanet atmosphere. Detailed models for specific planets should be developed on case-by-case basis and validated by comparison with relevant observations. Instead, we use the simulations to explore the validity of the above simplifications and highlight the importance of the extent of the lower and middle atmosphere to atmospheric escape from many close-in exoplanets. 

The model solves multi-species equations of continuity with bulk flow equations of momentum and energy in the radial direction. It includes photoionization by stellar XUV radiation, thermal ionization, recombination, charge exchange, advection and thermal escape of ions and neutrals, diffusion, viscous drag, heating by photoionization, adiabatic heating and cooling, heat conduction, viscous dissipation, and radiative cooling by recombination, the emission of Lyman~$\alpha$ photons and H$_3^+$ infrared emissions. The model is coupled to the KPP model for robust numerical simulations of chemistry \citep{sandu06}. In this work, we include the species H, He, H$_2$, H$^+$, He$^+$, H$_2^+$, H$_3^+$, HeH$^+$, and electrons. The relevant chemical reactions as well as the details of the heating and cooling parameterizations are given in Appendix~\ref{ap:atmosmodel}.   

The model uses a stretched grid with a spacing $\Delta r_0 =$~10 km at the lower boundary and a stretch factor of 1.014. Numerical stability of the time integration is preserved by the use of semi-implicit diffusion \citep{jacobson99} and van Leer flux-limited advection \citep{vanleer79}. The lower boundary is at $p_0 =$~10$^{-6}$~bar where the lower boundary conditions for temperature, radius, and species abundances are obtained from our lower and middle atmosphere models (Section~\ref{subsc:lowatm}). The outflow velocity at the lower boundary is based on mass conservation. The upper boundary conditions depend on the application. If the exobase is below the sonic point, the model assumes that individual species escape at a modified Jeans rate that accounts for ionization and Roche potential effects \citep{koskinen13a,koskinen13b,koskinen15}. If the exobase is above the sonic point, escape is hydrodynamic and the model extrapolates upper boundary conditions for the outflow velocity, species densities, and temperature at a constant slope \citep{tian05}. The latter applies to all simulations in this work. 

\section{Results}
\label{sc:results}
 
\subsection{A fundamental stability limit}
\label{subsc:overflow2}

\noindent
Once the atmosphere of a planet reaches the Roche lobe, the mass loss rate rapidly increases. Based on this, we define a stability limit for gaseous envelopes by using the simple theory from Sections~\ref{subsc:lowatm} and \ref{subsc:overflow1}, and relate the limit to the equilibrium shape of a planet near the surface pressure level. Here, we ignore the heating of the upper atmosphere and assume instead that the effective temperature of the planet applies throughout the atmosphere up to the exobase or the Roche lobe. This produces a lower limit on mass loss and conveniently, the results depend only on $M_p$, $R_p$, $M_s$, $a$, stellar bolometric luminosity $L_s$, and the Bond albedo of the planet that we set to $A_B =$~0.3. We focus on sun-like host stars, so that $M_s = M_{\sun}$ and $L_s = L_{\sun}$. In order to highlight the underlying physics that determines the stability limit, we consider a tidally locked Jupiter-like planet orbiting a sun-like star ($M_p =$~1898.19~$\times$~10$^{24}$~kg, $R_p =$~71,492 km) to represent hot Jupiters and a tidally locked Uranus-like planet orbiting a sun-like star ($M_p =$~86.813~$\times$~10$^{24}$~kg, $R_p =$~25,559 km) to represent hot Neptunes. We assume that the polar radii of the planets at 1 bar are equal to the radii of Jupiter and Uranus. 

\begin{figure}
  \epsscale{1.1}
  \plotone{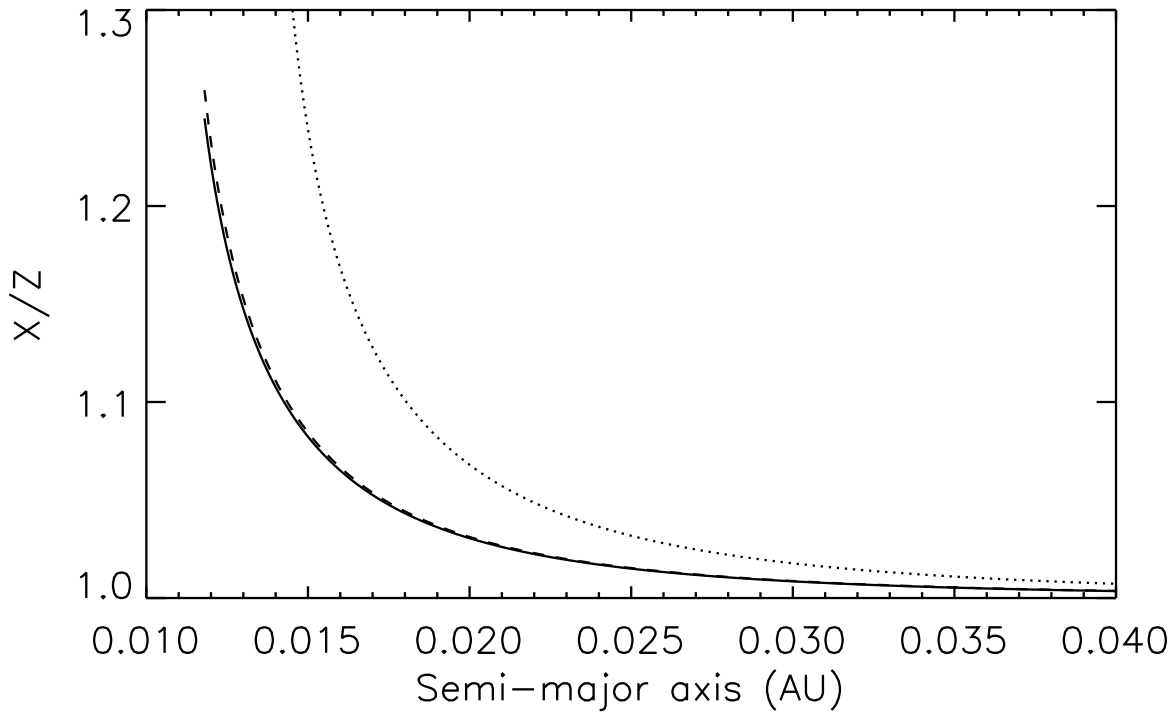}
  \caption{The ratio of the surface substellar radius $X$ to the polar radius $Z$ for Uranus-like (solid line), Jupiter-like (dashed line), and Saturn-like (dotted line) planets orbiting a sun-like star.}
  \label{fig:urratio}
\end{figure}    

The shape of the 1 bar surface changes with decreasing orbital distance as the planets become increasingly distorted by the stellar tide. Figure~\ref{fig:urratio} shows the ratio of the substellar radius to the polar radius, $X/Z$, for Uranus, Saturn, and Jupiter-like planets as a function of orbital distance. Here, the capital letters in the $X/Z$ ratio refer to the observable surface of the planets. We use lower case letters for the perturbation of the pressure levels generally. The results are similar for Jupiter and Uranus-like planets because they have similar bulk densities. The ratio is close to unity at orbital distances of $a \gtrsim$~0.03 AU ($P \gtrsim$~1.9 days). At orbital distances of $a \lesssim$~0.02 AU ($P \lesssim$~1.03 days), the ratio increases rapidly with decreasing orbital distance. Saturn has a lower bulk density and therefore, the ratio begins to increase at somewhat wider orbits. The results are consistent with the classical Roche limit for tidal destruction of a fluid planet:
\begin{equation}
R_L \approx 2.455 R_s \left( \frac{\rho_s}{\rho_p} \right)^{\frac{1}{3}}
\end{equation}
where $\rho_s$ and $\rho_p$ are the bulk densities of the star and the planet, respectively. Hereafter, we focus on the Jupiter and Uranus-like planets as an example. In Section~\ref{sc:discussion}, we generalize our approach to different planets. For both Jupiter and Uranus, the Roche limit is $R_L \approx$~0.0118 AU ($P =$~0.47 days). In Figure~\ref{fig:urratio}, the $X/Z$ ratio exceeds 1.2 at $a \lesssim$~0.0124 AU ($P \lesssim$~0.5 days), as the planets approach their Roche limits. We note that $X/Z \approx$~1.5 once the surface of the planet fills the Roche lobe.

As tidal distortion increases, the upper atmosphere reaches the Roche lobe first. To illustrate this, we use the approach from Section~\ref{subsc:lowatm} to calculate the extent of the atmosphere for the planets. We remind the reader that we ignore the heating of the upper atmosphere by stellar XUV radiation and photochemistry here. Stellar XUV heating would produce a hotter upper atmosphere and photochemistry would lead to more efficient ionization and dissociation of molecules, both of which increase the extent of the atmosphere and mass loss rate. Thus, the results here represent lower limits on both. 

\begin{figure}
  \epsscale{1.1}
  \plotone{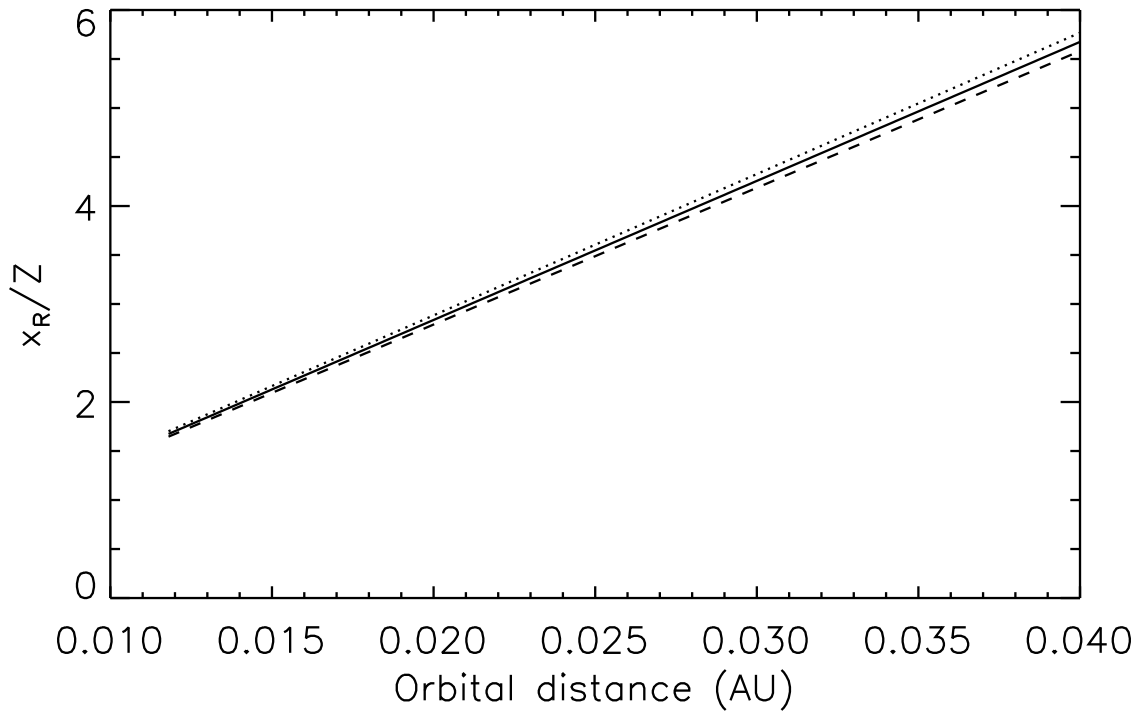}
  \caption{Radial distance $x_R$ to the L1 point in units of the polar 1 bar radius $Z$ for Uranus-like (solid line) and Jupiter-like (dashed line) planets orbiting a sun-like star. The dotted line shows $x_R/R_p$ for a Uranus-like planet based on the approximate expression from \citet{erkaev07} (equation~\ref{eqn:erkaevl1}).}
  \label{fig:top_atmh}
\end{figure}  

\begin{figure}
  \epsscale{1.1}
  \plotone{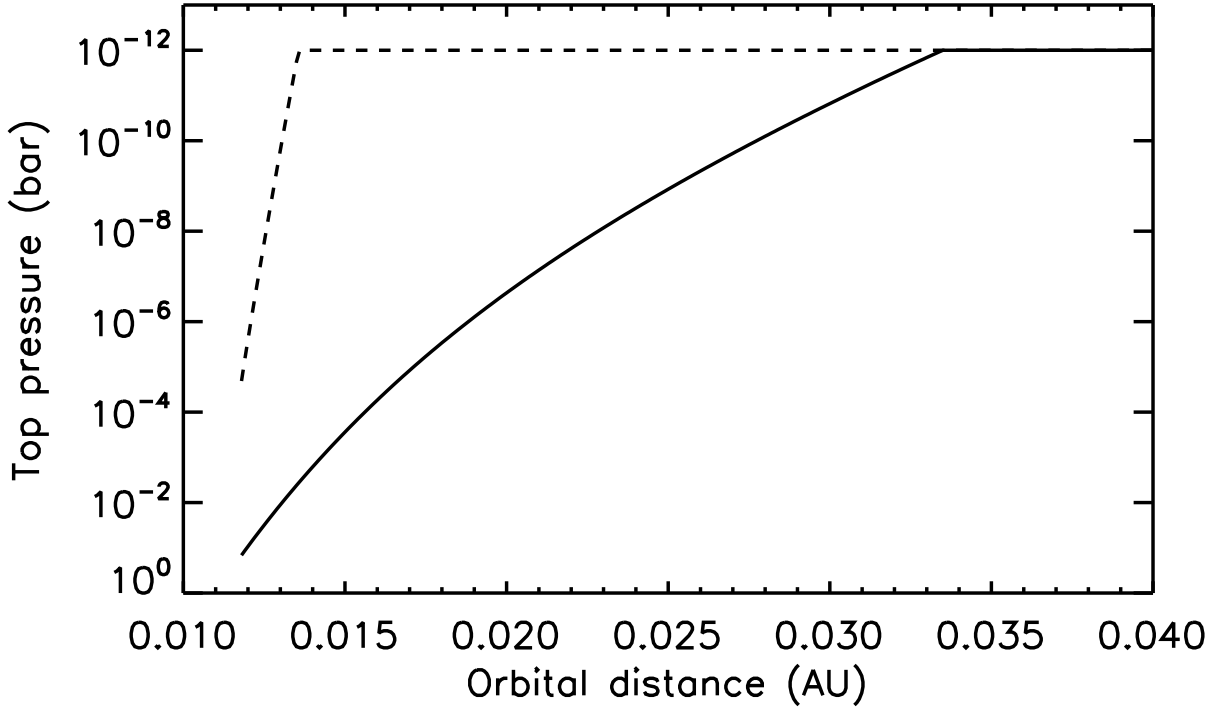}
  \caption{Pressure at the top of the atmosphere for Uranus-like (solid) and Jupiter-like (dashed) planets orbiting a sun-like star. The top of the atmosphere is either the exobase ($p =$~10$^{-12}$ bar) or the Roche lobe when the latter is below the exobase.}
  \label{fig:top_atm}
\end{figure}

We integrate equation~(\ref{eqn:preslevels}) numerically at the substellar point by using a pressure grid that ranges from $p_0 =$~1 bar to the exobase at $p =$~10$^{-12}$~bar with a pressure level spacing of 0.01 scale heights. The top of the atmosphere is at $p =$~10$^{-12}$~bar if the L1 point is above this level. If not, the top of the atmosphere is the L1 point. Figure~\ref{fig:top_atmh} shows the radius of the L1 point for Uranus and Jupiter-like planets orbiting a sun-like star. It also shows the radius at the L1 point for a Uranus-like planet based on the approximate expression from \citet{erkaev07} (equation~\ref{eqn:erkaevl1}). Figure~\ref{fig:top_atm} shows the corresponding pressure at the top of the atmosphere. By our definition, once the Roche lobe reaches down to the exobase, the atmosphere undergoes Roche lobe overflow and the pressure at the top of the atmosphere (L1 point) exceeds 10$^{-12}$ bar (see Section~\ref{subsc:overflow1}). A Uranus-like planet has lower gravity and a more extended atmosphere than a Jupiter-like planet. Thus, a Uranus-like planet undergoes Roche lobe overflow at $a \lesssim$~0.033 AU ($P \lesssim$~2.2 days) while a Jupiter-like planet undergoes Roche lobe overflow at $a \lesssim$~0.014 AU ($P \lesssim$~0.6 days). 

Many studies of exoplanet evolution rely on simple energy-limited escape to represent stellar XUV-driven photoevaporation \citep[e.g.,][]{owen20}. Previous work has shown that energy-limited escape can substantially overestimate mass loss unless the atmosphere undergoes hydrodynamic escape and adiabatic cooling due to escape mostly balances heating \citep{koskinen14}. Here, we focus on the opposite end of escape regimes. Once the middle atmosphere reaches the Roche lobe, the atmosphere effectively enters a `low gravity' regime and is free to escape \citep[see also][]{jackson17,kubyshkina18}. This regime cannot be captured by adding a simple correction to energy-limited escape that represents the potential difference between the surface of the planet and the L1 point \citep[][equation \ref{eqn:erkaevdu} here]{erkaev07}. Even with this correction, energy-limited escape substantially underestimates the mass loss rate due to Roche lobe overflow.  

\begin{figure}
  \epsscale{1.1}
  \plotone{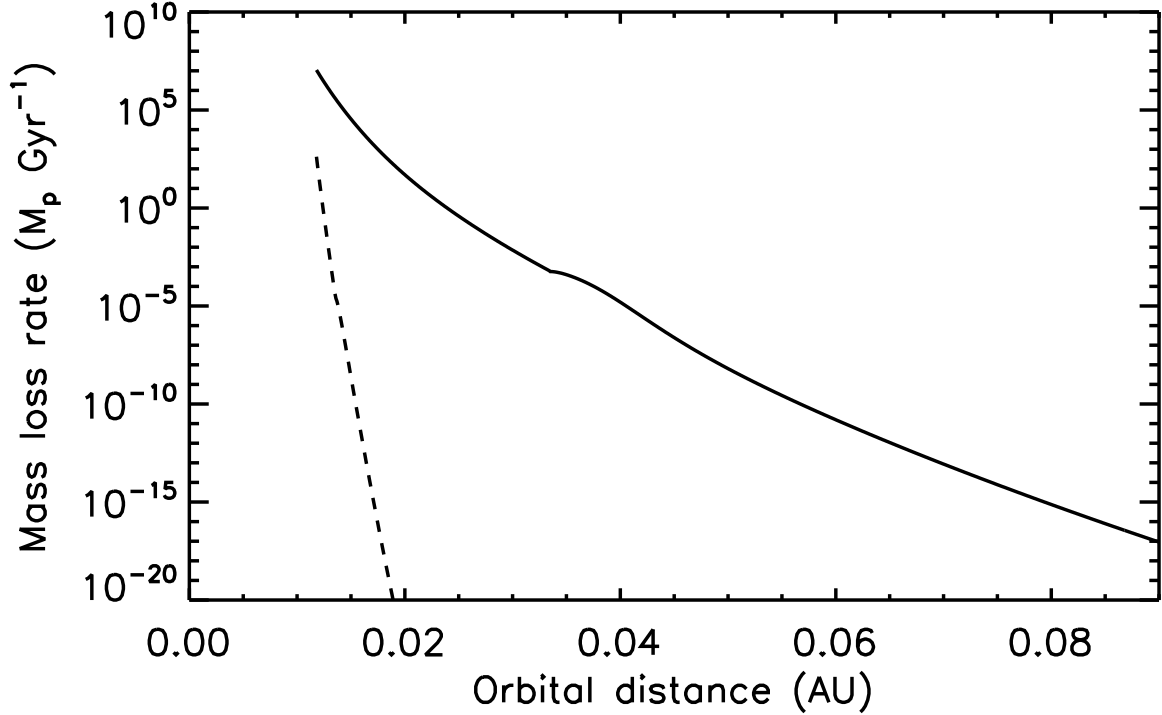}
  \caption{Minimum mass loss rate for Uranus-like (solid line) and Jupiter-like (dashed line) planets orbiting a sun-like star, based on Jeans escape or Roche lobe overflow due to the heating of the lower and middle atmosphere.}
  \label{fig:mloss1}
\end{figure}

\begin{figure}
  \epsscale{1.1}
  \plotone{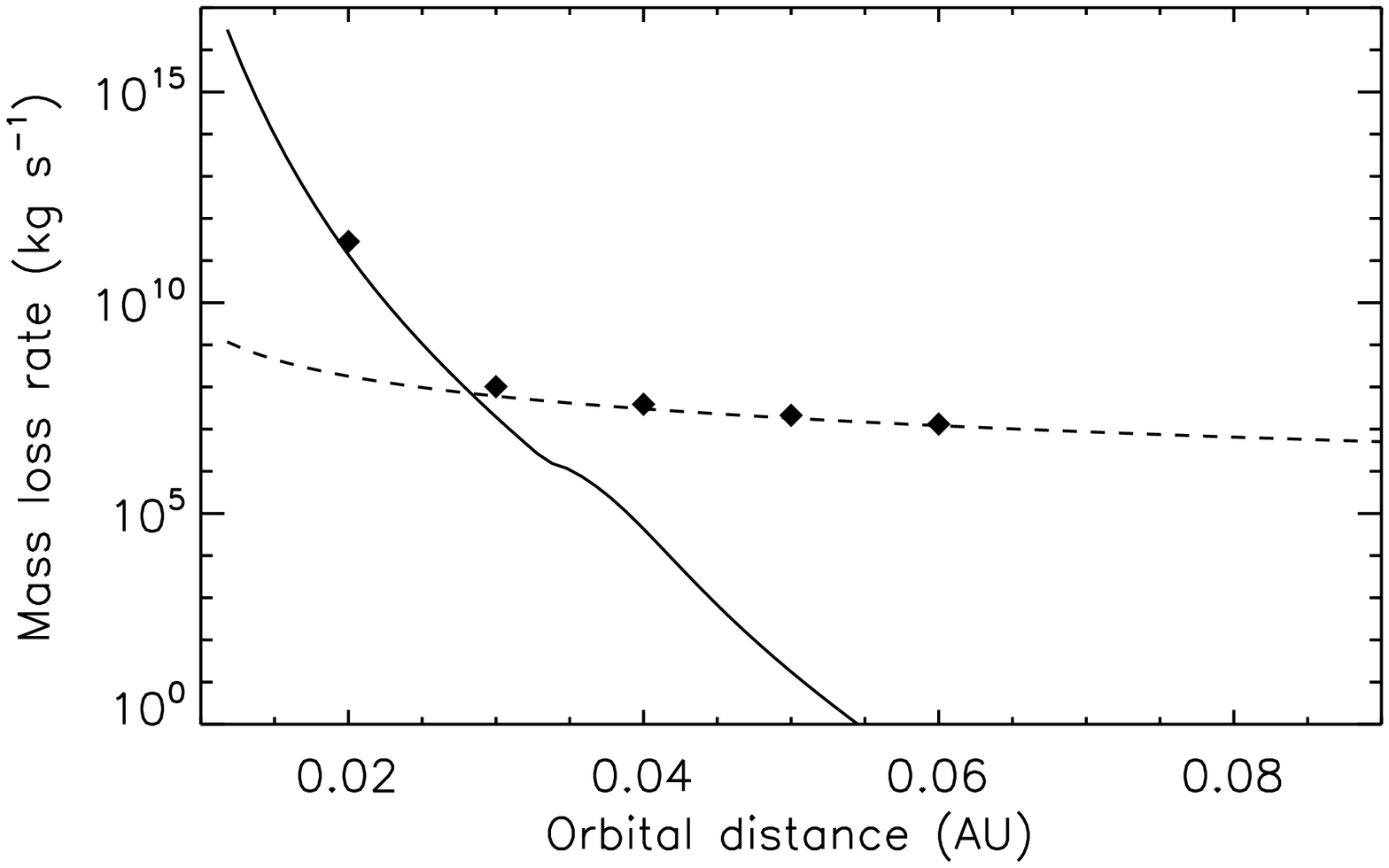}
  \caption{Mass loss rate for a Uranus-like planet orbiting a sun-like star based on Roche lobe overflow (solid line) and stellar XUV-driven energy-limited escape \citep[dashed line,][]{erkaev07}. The diamonds show the mass loss rates calculated by our detailed numerical model of hydrodynamic escape (Section~\ref{subsc:escape_model}).}
  \label{fig:mloss2}
\end{figure}

We calculate the corresponding mass loss rates due to Roche lobe overflow by using equations~(\ref{eqn:mloss1})--(\ref{eqn:mloss2}). As we note above, these are lower limits on mass loss since we ignore the heating of the upper atmosphere entirely. Figure~\ref{fig:mloss1} shows the mass loss rate at different orbital distances in units of $M_p$~Gyr$^{-1}$ while Figure~\ref{fig:mloss2} compares it with the mass loss rate based on energy-limited photoevaporation (equations~\ref{eqn:enerlim} and \ref{eqn:erkaevdu} with $r_e = R_p$). For energy-limited escape, we use the mean solar XUV flux of 1.6 W~m$^{-2}$ at 0.05 AU integrated over the wavelength range of 0.1-91.1 nm \citep{ribas05}, with an efficiency factor of $\epsilon =$~1. As we note earlier, energy-limited escape rates for Uranus and Jupiter-like planets are similar because the bulk densities of the planets are similar. 

As indicated by Figures~\ref{fig:top_atm} and~\ref{fig:mloss2}, stellar XUV heating of the upper atmosphere is the dominant driver of atmospheric escape at orbital distances where the Roche lobe is above the exobase. At short orbital distances where the Roche lobe is below the exobase, however, Roche lobe overflow dominates over photoevaporation. Under these circumstances, energy-limited escape significantly underestimates the mass loss rate because it does not account for the extent of the atmosphere. In this case, escape is effectively powered by the bolometric luminosity of the star, which is more than five orders of magnitude higher than the XUV luminosity. Tidal heating of the planet can provide an additional energy source \citep{li10} but is not necessarily required.

For a Uranus-like planet orbiting a sun-like star, the cumulative mass lost in 1 Gyr exceeds the mass of the planet at $a \lesssim$~0.024 AU ($P \lesssim$~1.36 days). At this limiting distance, the pressure at the Roche lobe is $p_R =$~3.3~$\times$~10$^{-9}$ bar and $X/Z =$~1.017. The same limits for complete mass loss in only 100 Myr are $a \lesssim$~0.022 AU ($P \lesssim$~1.2 days), $p_R =$~4~$\times$~10$^{-8}$ bar, and $X/Z =$~1.024. For a Jupiter-like planet, the limits for complete mass loss in 1 Gyr are $a \lesssim$~0.0124 AU ($P \lesssim$~0.5 days), $p_R =$~4.6~$\times$~10$^{-8}$ bar, and $X/Z =$~1.19. The same limits for complete mass loss in only 100 Myr are $a \lesssim$~0.0122 AU ($P \lesssim$~0.5 days), $p_R =$~4.8~$\times$~10$^{-7}$ bar, and $X/Z =$~1.21. In other words, Jovian planets experience tidal destruction once the surface $X/Z$ ratio exceeds 1.2 near the classical Roche limit. The limiting $X/Z$ ratio for sub-Jovian planets, however, is smaller and they experience envelope loss outside of the Roche limit. Thus, the visible surface of a planet need not fill the Roche lobe for the planet to lose its envelope over a relatively short time. Our results show that rapid mass loss occurs once the 10$^{-8}$--10$^{-7}$ bar level in the upper atmosphere reaches the Roche lobe. We note that the results here do not account for the change in mass loss once the envelope is substantially depleted and instead assume that the mass loss rate is constant over time (see Section~\ref{sc:discussion} for related discussion).

\subsection{Numerical modeling}
\label{subsc:num_model}

\noindent
In order to test the validity of the simplifying assumptions above, we use a multi-species model of hydrodynamic escape (see Section~\ref{subsc:escape_model} and Appendix~\ref{ap:atmosmodel}) to simulate the transition from stellar XUV-driven escape to Roche lobe overflow in more detail. Obviously, this model accounts for the heating of the upper atmosphere by stellar XUV radiation, unlike the results presented above. We focus on a Uranus-like planet orbiting a sun-like star. We calibrate the model at 0.05 AU where Roche lobe overflow is not expected to be significant. This provides a direct comparison to models of the prototypical hot Jupiter HD209458b that has a similar orbital distance around a sun-like star \citep{koskinen13a,koskinen13b}. Our reference Model A does not include Roche lobe overflow. In Model B, we replace spherically symmetric gravity with the substellar gravity based on the Roche potential (equation~\ref{eqn:ssgrav}). This corresponds to escape through the L1 point. In both cases, we use the same mean solar XUV spectrum as \citet{koskinen13a,koskinen13b} and assume uniform redistribution of energy around the planet, in line with energy-limited escape. 

The radial mass flux constant predicted by Model A at $a =$~0.05 AU is $F_M = \rho v r^2 =$~1.5~$\times$~10$^6$ kg~s$^{-1}$~sr$^{-1}$. Multiplying this by a solid angle factor $4 \pi$ gives a global mass loss rate of $\dot{M} =$~1.9~$\times$~10$^7$ kg~s$^{-1}$. \citet{owen20} compared 3D simulations of the upper atmosphere of the warm Neptune GJ436b \citep{shaikhislamov18} with 1D globally averaged simulations of the same planet by our model \citep{parkeloyd17} and found good agreement on the mass loss rates. This means that our 1D model can be used to calculate global mass loss rates, and that atmospheric extent and the details of the energy balance in the upper atmosphere are likely more important for predicting accurate mass loss rates than adding multiple dimensions to the model (see Section~\ref{sc:discussion} for related discussion).

\begin{figure}
  \epsscale{1.18}
  \plotone{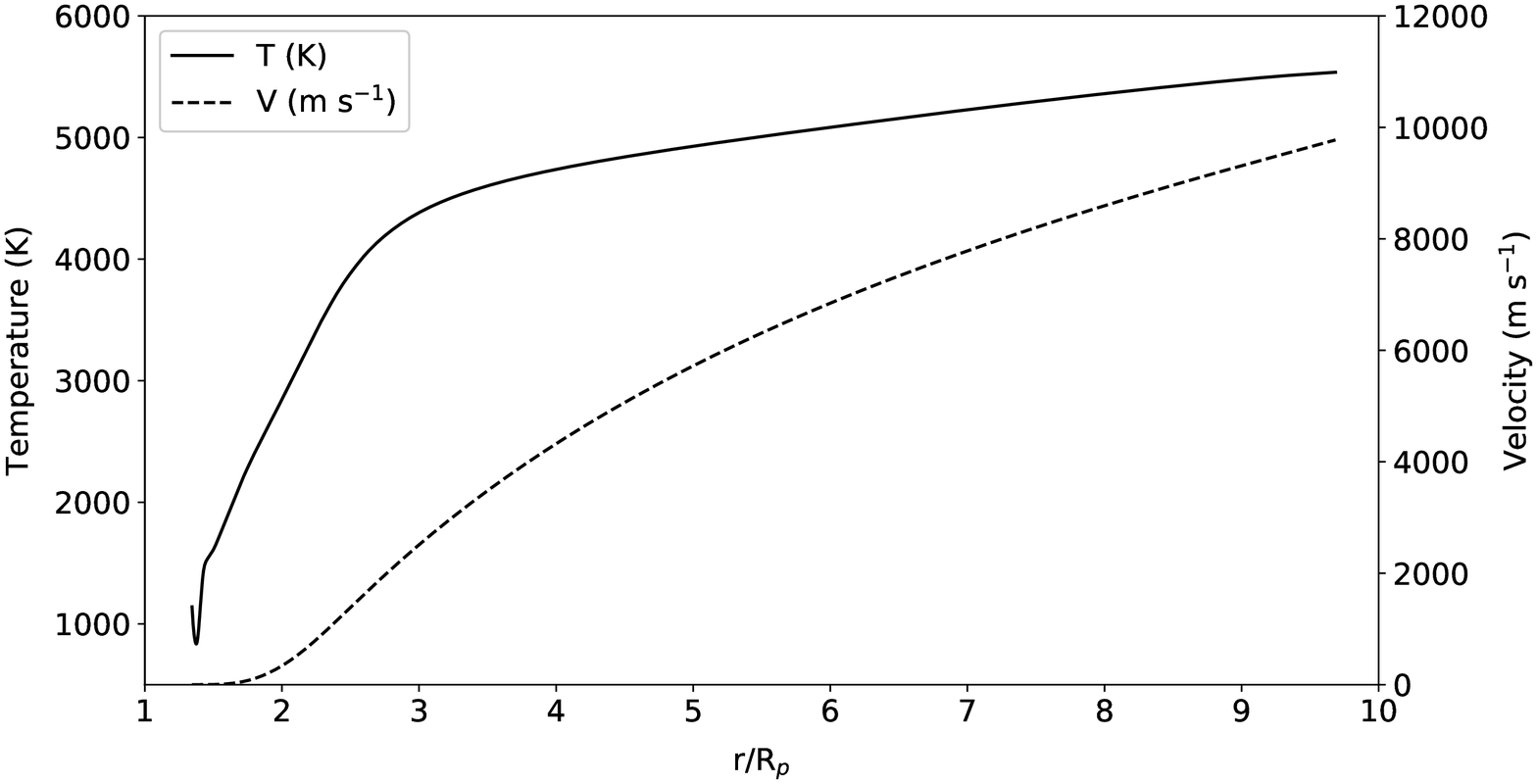}
  \caption{Temperature and outflow velocity predicted by our model for a Uranus-like planet orbiting a sun-like star at the orbital distance of 0.05 AU.}
  \label{fig:tempvelA}
\end{figure}

Figure~\ref{fig:tempvelA} shows the temperature and outflow velocity profiles for Model A at $a =$~0.05 AU. The lower boundary of the model is at $r_0 =$~1.34 $R_p$ ($p_0 =$~10$^{-6}$ bar, $T_0 =$~1,140 K) where the boundary conditions are consistent with our models of the lower and middle atmosphere (see Section~\ref{subsc:lowatm}). Above the lower boundary, the temperature first slightly decreases with radius, and then increases rapidly to about 4,400~K around $r \approx$~3 $R_p$. At higher radii, the temperature increases slowly with radius to about 5,540~K at the upper boundary of the model. The outflow velocity in the model is significant at $r \gtrsim$~2~$R_p$ and it increases slowly with radius to about 9.8 km~s$^{-1}$ at the upper boundary. The adiabatic sound speed ranges from about 2.5 km~s$^{-1}$ at the lower boundary to about 8 km~s$^{-1}$ at the upper boundary. The corresponding range of values for the isothermal sound speed is from about 2 km~s$^{-1}$ to 6 km~s$^{-1}$. The adiabatic and isothermal sonic points are at 6.7 $R_p$ and 4.7 $R_p$, respectively, while the L1 point is at 7.1 $R_p$. The exobase is above the upper boundary of the model and therefore, the atmosphere escapes hydrodynamically. 

\begin{figure}
  \epsscale{1.3}
  \plotone{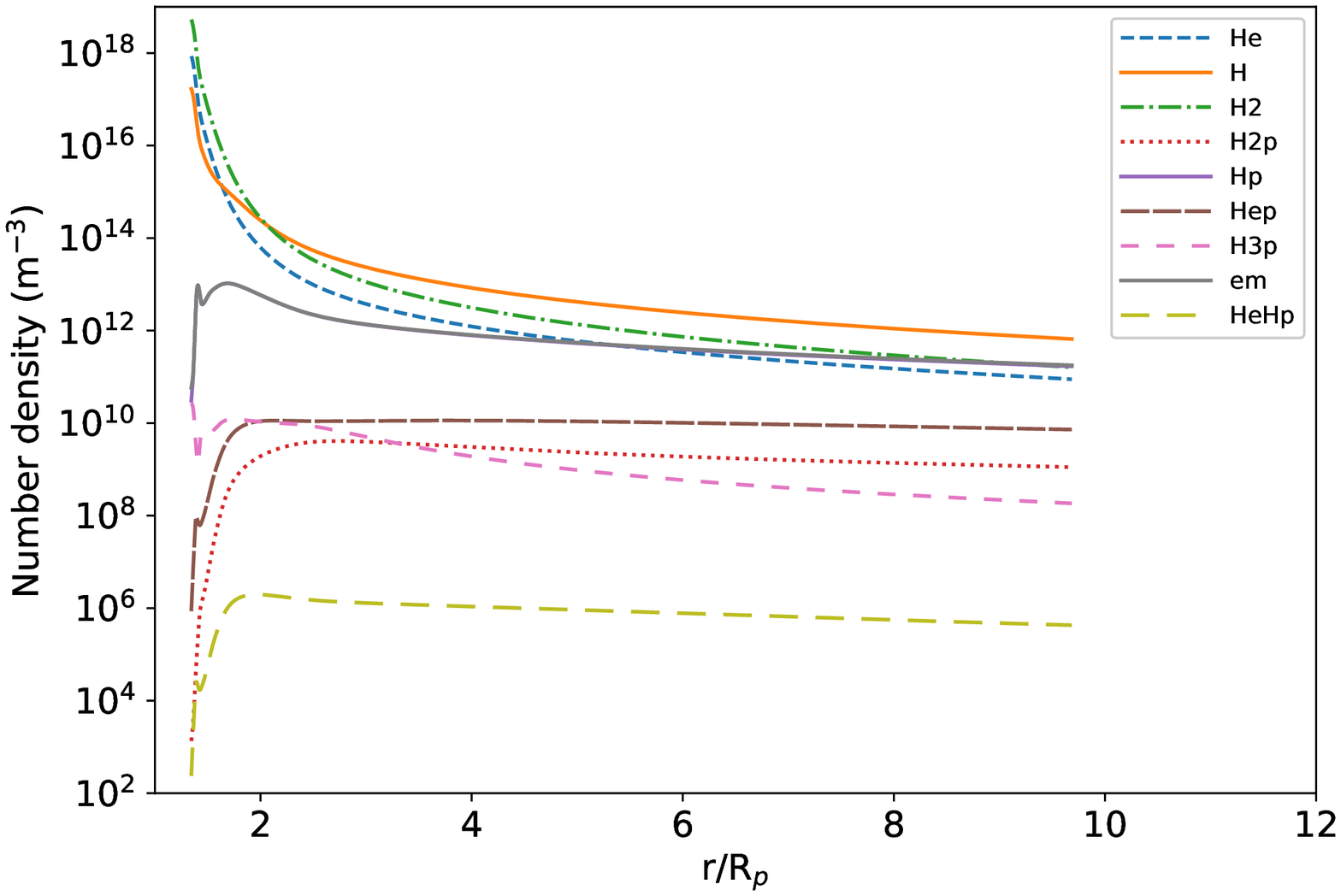}
  \caption{Composition predicted for a Uranus-like planet orbiting a sun-like star at the orbital distance of 0.05 AU. Note that the H$^+$ profile is practically identical to the electron (em) density profile.}
  \label{fig:compoA}
\end{figure}

The composition is shown by Figure~\ref{fig:compoA}. The volume mixing ratio of H$_2$ at the lower boundary is about $q_0 =$~0.84 while H is a minor species with $q_0 =$~0.026. H becomes the dominant species at around $r \approx$~2 $R_p$ but the density of H$_2$ remains significant at all altitudes. The dominant ion is H$^+$, with a density that is at least two to three orders of magnitude higher than the densities of other ions. Charge balance means that the electron density is equal to the proton density. The density of neutral H is higher than the proton density at all radii. 

Simple ionization $H_2 + h\nu \rightarrow H_2^+ + e$ is the dominant photoionization channel for H$_2$, and the resulting H$_2^+$ reacts quickly with H$_2$ to form H$_3^+$ (see Appendix~\ref{ap:atmosmodel}). Dissociative recombination of H$_3^+$ is a rapid process that keeps the density of H$_3^+$ relatively small. Despite its small abundance, infrared emissions by H$_3^+$ are nevertheless significant. These emissions are regularly detected on solar system giant planets where they are used to probe the ionosphere with remote observations. On Jupiter, they also act as an important coolant \citep{miller13}. The density of the other simple molecular ion, HeH$^+$, created by the reaction $He^+ + H_2 \rightarrow HeH^+ + H$, is also significant in our model. Rovibrational emissions by HeH$^+$ have been detected recently from planetary nebulae, constituting the first detection of the ion in astrophysical environments \citep{gusten19,neufeld20}. 

\begin{figure}
  \epsscale{1.3}
  \plotone{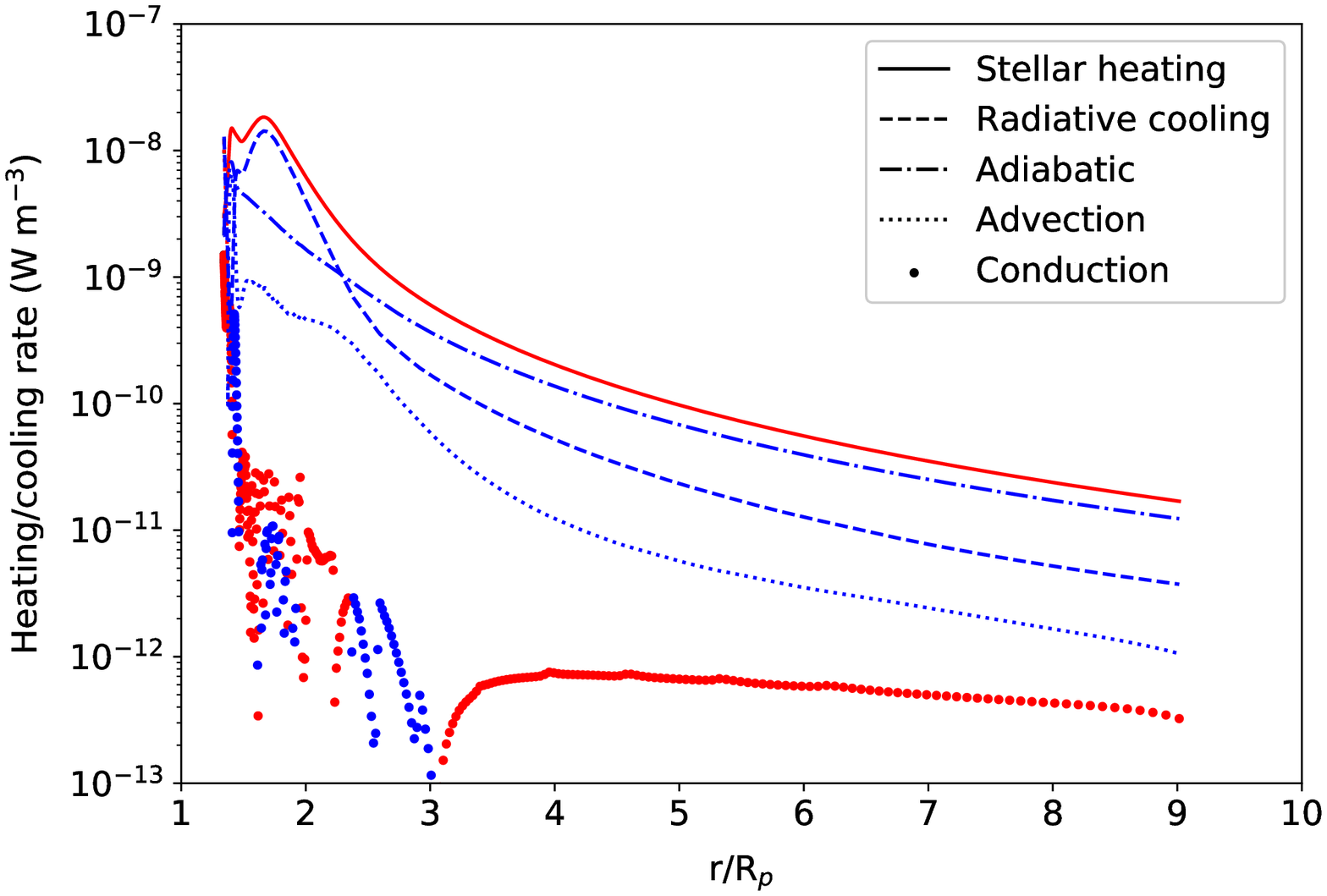}
  \caption{Heating (red) and cooling (blue) rates for a model of a Uranus-like planet orbiting a sun-like star at the orbital distance of 0.05 AU.}
  \label{fig:enerbA}
\end{figure}

The stellar XUV heating peak is at a radius of 1.66 $R_p$ ($p =$~2.1~$\times$~10$^{-9}$ bar), as demonstrated by Figure~\ref{fig:enerbA} that shows the prominent heating and cooling terms. The peak volume heating rate is 1.8~$\times$~10$^{-8}$ W~m$^{-3}$. At $r \lesssim$~2.3 $R_p$, heating is mostly balanced by H$_3^+$ cooling. At higher radii, the outflow velocity is significant and adiabatic cooling due to the expansion of the atmosphere primarily balances the stellar heating rate, with a smaller contribution from H$_3^+$ cooling. Advection makes a small contribution to cooling and heat conduction is largely negligible. We note that the rate for heat conduction in Figure~\ref{fig:enerbA} is diagnostic only. In practice, the model uses a semi-implicit method to update the temperature profile for the effect of conduction and the diagnostic rate is calculated \textit{ex post facto} with finite difference methods that produce numerical noise at small values.  

We now compare our results with energy-limited escape to find values of $\epsilon$ and $\Delta U$ in equation~(\ref{eqn:enerlim}) that agree with our detailed simulations. With a stellar XUV luminosity of $L_{\text{XUV}} =$~1.12~$\times$~10$^{21}$ W, the flux incident on Model A at $a =$~0.05 AU is $F_{\text{XUV}} =$~1.6 W~m$^{-2}$. Thus, the globally averaged flux incident on the planet is about 0.4 W~m$^{-2}$. The effective flux that heats the upper atmosphere can be obtained from our model as
\begin{equation}
\frac{\epsilon F_{\text{XUV}}}{4} = \int_{r_0}^{\infty} \left[ H(r) - C(r) \right] \text{d} r = 0.11 \ \text{W}~\text{m}^{-2}
\end{equation}  
where $H(r)$ and $C(r)$ are the radiative volume heating and cooling rates, respectively, from the model. Therefore, the model suggests a mass loss efficiency $\epsilon =$~0.28 for energy-limited escape. With this efficiency, equation~(\ref{eqn:enerlim2}) gives a mass loss rate of 4~$\times$~10$^6$ kg~s$^{-1}$, which is 4.8 times lower than the actual mass loss rate predicted by Model A. The model shows, however, that the heating peak is at $r_e = 1.66 R_p$. Replacing $R_p$ in equation~(\ref{eqn:enerlim2}) with the radius $r_e$ at the heating peak yields an energy-limited escape rate of 1.8~$\times$~10$^7$ kg~s$^{-1}$ that agrees roughly with Model A.

In this case, neglecting the extent of the atmosphere underestimates the energy-limited mass loss rate by a factor of five. This factor is larger than similar factors derived for typical hot Jupiters because of the larger extent of the envelopes of Neptune-mass planets. \citet{owen17} suggested that the effect of the `expansion radius' can be folded into the efficiency factor. This is not an ideal solution for low-mass planets with H/He envelopes because our results imply efficiencies close to unity or even higher that are unlikely to be adopted by modelers. In fact, the escape rate of 1.8~$\times$~10$^7$ kg~s$^{-1}$ is comparable to the energy-limited escape rate in Figure~\ref{fig:mloss2} that we obtained by using equations~(\ref{eqn:enerlim}) and (\ref{eqn:erkaevdu}) with 100\% efficiency and $r_e = R_p$.      

Alternatively, the base radius for escape can be estimated from hydrostatic equilibrium as \citep[e.g.,][]{lopez17}
\begin{equation}
R_b \approx R_p + H_1 \ln \left( \frac{p_1}{p_b} \right)
\end{equation}
where $H_1$ is the scale height based on $m_1 =$~2.3 amu and the effective temperature $T_1$ of the planet, $p_1 =$~1 bar is the surface pressure and $p_b =$~1 nbar is the base pressure for escape. At $a =$~0.05 AU, we obtain $H =$~470 km at the 1 bar level and $R_b \approx$~1.38 $R_p$ at the 1 nbar level. A better approximation can be obtained by including the dependency of gravity on radius in the above equation:
\begin{equation}
\Gamma (R_b) \approx \Gamma_1 + \ln \left( \frac{p_b}{p_1} \right)
\end{equation}   
where 
\begin{equation}
\Gamma = \frac{G M_p m_1}{k T_1 r}
\end{equation}
is the thermal escape parameter. We obtain $\Gamma_0 =$~54.6 and $\Gamma =$~33.9 where the latter corresponds to $R_b \approx$~1.6 $R_p$. This is a straightforward way of roughly reproducing the mass loss rate predicted by our detailed model with $\epsilon =$~0.3. It is reasonably accurate even though it does not account for all the complex physics included in our model. 

Model B represents escape through the L1 point. The radial mass flux based on the model is $F_M =$~1.7~$\times$~10$^6$ kg~s$^{-1}$~sr$^{-1}$. This is higher by a factor of 1.13  than the radial mass flux predicted by Model A. The outflow velocity in Model B is somewhat higher than in Model A, reaching a velocity of 12 km~s$^{-1}$ at the upper boundary. Due to the increased mass loss rate and the strength of adiabatic cooling, the temperature of 4740 K at the upper boundary is lower than in Model A. Otherwise, Models A and B are qualitatively similar. The adiabatic and isothermal sonic points in Model B are at 5.2 $R_p$ and 4.2 $R_p$, respectively. The sonic points are below the L1 point of 7.1 $R_p$, and the equipotential surfaces that coincide with them are approximately spherical. Therefore, the atmosphere escapes through the sonic surface at a global mass loss rate of $4 \pi F_M =$~2.1~$\times$~10$^7$ kg~s$^{-1}$.

\subsubsection{Comparison with hot Jupiter simulations}

\noindent
In order to highlight the main qualitative differences between escape models for hot Jupiters and our model for a Uranus-like planet, we briefly describe the key differences between the model above and the simulations of the prototype hot Jupiter HD209458b that also orbits a sun-like star at a distance of about 0.05 AU. We predict temperatures of 4000--5000 K for the hot Uranus that are significantly cooler than the peak temperature of 8,000--12,000 K predicted for HD209458b \citep{koskinen13a,koskinen13b}. In addition, the temperature profile for hot Uranus lacks a well defined peak that appears in hot Jupiter simulations. Both of these differences are explained by the lower gravity of the hot Uranus model that makes it more susceptible to the effects of atmospheric escape that acts as the primary cooling mechanism above the heating peak.  

Models of HD209458b predict significant dissociation of H$_2$ near the bottom of the thermosphere that removes practically all H$_2$ molecules and related molecular ions from the outflow \citep{yelle04,koskinen13a}. In contrast to this, the density of H$_2$ is significant at all radii of the hot Uranus model. This leads to the appearance of the molecular ions H$_3^+$ and HeH$^+$ at high altitudes, neither of which are significantly produced in hot Jupiter simulations. Another key difference between the composition of the hot Uranus model and the hot Jupiter models is that the density of neutral H is higher than the proton density at all radii. The prevalence of neutral H and H$_2$ at high radii in the hot Uranus model is due to a combination of a relatively low temperature and high escape flux that replenishes H$_2$ and H at high altitudes where they would otherwise be predominantly dissociated and ionized (see Appendix~\ref{ap:atmosmodel}).   

\subsubsection{Transition to Roche lobe overflow}

\noindent
Here, we present results from Model B at orbital distances of $a =$~0.06 AU, 0.04 AU, 0.03 AU, and 0.02 AU. Figure~\ref{fig:mloss2} compares the global mass loss rates predicted by the model for each orbital distance with the minimum mass loss rates based on the simple treatments of Roche lobe overflow (see Section~\ref{subsc:overflow2}) and energy-limited photoevaporation obtained for $\epsilon =$~1 with the Roche potential correction factor. For each model, the lower boundary conditions are consistent with the structure and composition of the lower and middle atmosphere (see Section~\ref{subsc:lowatm}). We note that, as before, Model B applies at the substellar point. For each model, the tidal perturbation of the lower and middle atmosphere is properly reflected by the lower boundary radius that is consistent with the $x/z$ ratio based on the Roche potential at $p_0 =$~10$^{-6}$ bar.

At orbital distances of $a =$~0.04--0.06 AU, the mass loss rate is close to the energy-limited escape rate. As expected, the dependency of the mass loss rate on the orbital distance is close to $\sim 1/a^2$ at these orbital distances. A slight deviation from this trend is caused by the effect of the stellar tide that enhances the mass loss rate as the orbital distance decreases. At $a =$~0.03 AU, the mass loss rate is higher by a factor of 2.6 than at $a =$~0.04 AU. This factor is larger than the expected factor of 1.8 increase based on a $1/a^2$ dependency, because Roche lobe overflow dominates over photoevaporation at $a \lesssim$~0.03 AU. At $a =$~0.02 AU, the global mass loss rate predicted by Model B is 2.8~$\times$~10$^{11}$ kg~s$^{-1}$. This value is twice as large as the minimum mass loss rate of 1.4~$\times$~10$^{11}$ kg~s$^{-1}$ due to Roche lobe overflow that we obtain in Section~\ref{subsc:overflow2}, while it is about three orders of magnitude higher than the energy-limited escape rate. We note that 2D simulations with a uniform redistribution of energy by \citet{guo13} indicate that globally averaged 1D hydrodynamic models with substellar gravity may overestimate Roche lobe overflow by a factor two. If this is true, our numerical model predicts a mass loss rate that agrees well with the simple theory that underlies equation~(\ref{eqn:mloss2}). This shows that (i) equation~(\ref{eqn:mloss2}) and related theory in Section~\ref{subsc:overflow1} provide a reasonably good approximation of mass loss due to Roche lobe overflow and (ii) photoevaporation does not significantly add to the mass loss rate once Roche lobe overflow sets in.

\begin{figure}
  \epsscale{1.18}
  \plotone{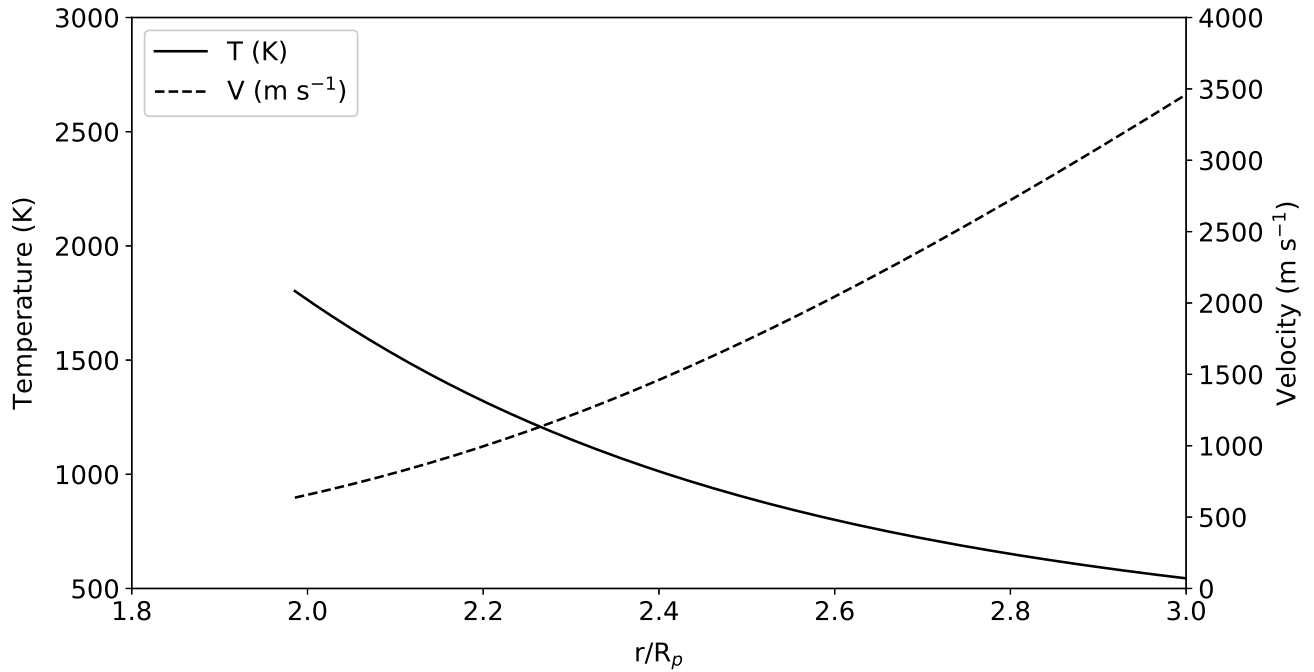}
  \caption{Temperature and outflow velocity predicted by our model for a Uranus-like planet orbiting a sun-like star at the orbital distance of 0.02 AU.}
  \label{fig:tempvel0_02}
\end{figure}

Figure~\ref{fig:tempvel0_02} shows the temperature and outflow profiles for Model B at $a =$~0.02 AU. The lower boundary radius at the substellar point is $r_0 =$~2.0 $R_p$ ($p_0 =$~10$^{-6}$ bar). Due to strong adiabatic cooling and advection based on the large escape rate, the temperature decreases with radius above the lower boundary to $T =$~650 K at the L1 point, which is located at $x_R =$~2.8 $R_p$. The adiabatic sonic point is at $r \approx$~2.7 $R_p$ and therefore, our numerical model supports the common approximation that the atmosphere escapes through the L1 point at the speed of sound \citep[e.g.,][]{li10,lai10}.   

\subsection{The shape of known transiting planets}
\label{subc:stability}

\noindent
We now explore the population of known transiting planets to make sure that their implied shape complies with the our new limit on the $X/Z$ ratio and identify the most extreme planets at risk of significant mass loss. Here, we focus only on transiting planets for which both the mass and radius have been measured. We include planets with orbital periods of less than 10 days ($a \approx$~0.09 AU around a sun-like host star) and masses higher than 10 $M_{\earth}$. For simplicity, we assume that the planets are tidally locked and have circular orbits. We obtain planet properties from the Extrasolar Planets Encyclopaedia (www.exoplanet.eu). We screen for confirmed planets with a primary transit detection, measured planet mass, and known stellar mass and luminosity. 

\begin{figure}
  \epsscale{1.1}
  \plotone{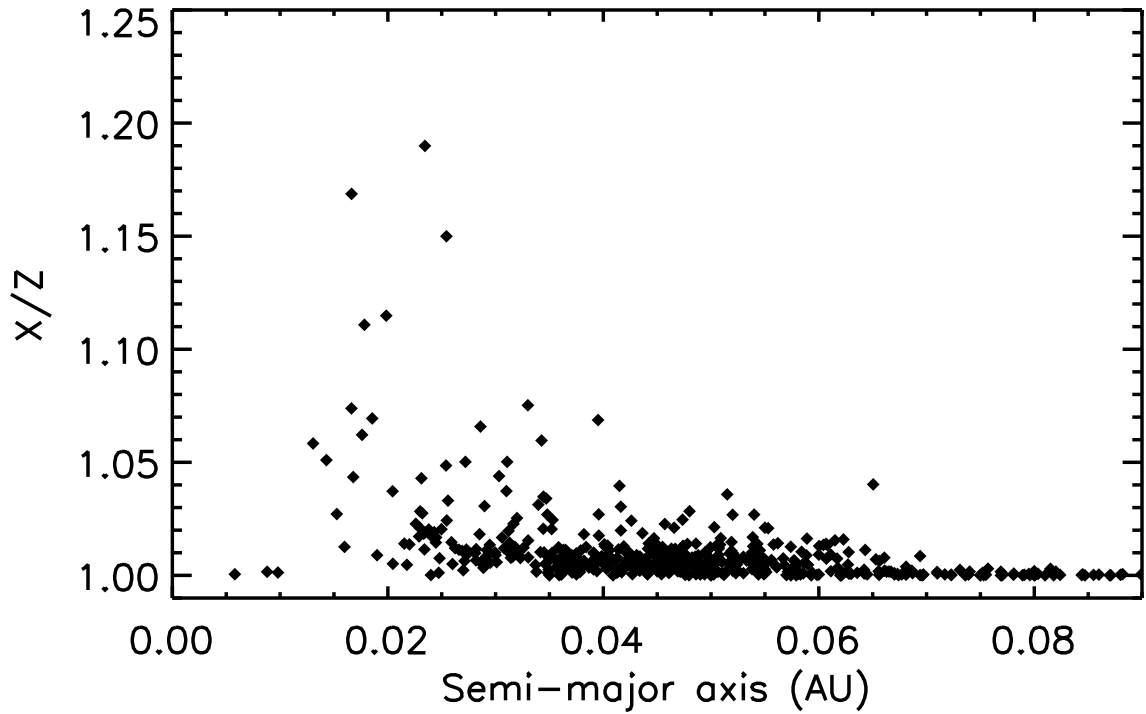}
  \caption{The ratio of the substellar radius $X$ to the polar radius $Z$ for transiting giant planets with known masses, radii, orbital distances, and host star masses.}
  \label{fig:eshape}
\end{figure}

We solve equation~(\ref{eqn:shape1}) numerically to calculate the implied ratio of the substellar radius to the polar radius $X/Z$ at the visible surface of the planets (i.e., the level probed by visual broadband transit). In all cases, we use Kepler's third law to calculate the orbital period and $\Omega$ based on the observed $M_p$, $M_s$, and $a$ instead of using the published orbital periods (see Section~\ref{subsc:periods}). The results are shown in Figure~\ref{fig:eshape}. The planets WASP-12b, WASP-19b, and WASP-121b have the highest $X/Z$ ratios of 1.19, 1.17, and 1.15, respectively. Two of these planets, WASP-12b and WASP-121b show evidence for the escape of multiple metals from their atmospheres, indicating large mass loss rates \citep{fossati10,sing19}. The mass loss rates of these planets exceed the stellar XUV-driven mass loss rate.

As expected, there are no transiting planets in this sample with implied $X/Z \gtrsim$~1.2, corresponding to a surface distortion that would lead to overflow of the atmosphere and rapid mass loss for Jovian planets. Furthermore, there are no transiting planets with masses between 10 $M_{\earth}$ and 20 $M_{\earth}$, similar to Uranus and Neptune, with $X/Z$ ratios higher than 1.06. The highest ratio of 1.06 in this mass range applies to the planet K2-266 b while the rest of the hot Neptunes have $X/Z$ ratios lower than 1.01. The ratio for K2-266 b is relatively high but it is also not particularly reliable. The uncertainty on planet's mass is relatively large and it could be twice as large as the listed mean value of 11.3 $M_{\earth}$ \citep{rodriguez18}. K2-266 b still has the highest $X/Z$ ratio if we increase the upper mass limit of the planets in the sample from 20 $M_{\earth}$ to 100 $M_{\earth}$ to include Saturn-like planets. This supports the results derived from the simple theory in Section~\ref{subsc:overflow2}. Surviving Neptune and Saturn-like planets have $X/Z$ ratios close to unity while all planets with higher $X/Z$ ratios are hot Jupiters.

\subsection{A note on the planet parameters}
\label{subsc:periods}

\noindent
In the previous section, we use $M_p$, $M_s$, and $a$ to calculate the orbital period and $\Omega$ based on Kepler's third law. This is because the values for the orbital period listed in both the exoplanet.eu catalogue and NASA Exoplanet Archive\footnote{https://exoplanetarchive.ipac.caltech.edu} can differ significantly from Kepler's third law. This problem is somewhat common; we found disagreements that exceed 5\% for about 11\% of the planets in our sample. In four cases, the listed combination of the orbital period, host star mass, and orbital distance would lead to unstable solutions for the shape of the planet based on the Roche potential that imply unrealistically high mass loss rates (see below). This is due to the interaction of the relevant parameters in equation~(\ref{eqn:rochepot}) when these parameters are inconsistent with a Keplerian orbit, a situation that is obviously unphysical. The deviations from a Keplerian orbit are due to uncertainties in the statistical fits to the transit and radial velocity data combined with, on occasion, the use of significantly inconsistent combinations of parameters. We note that the orbital period is measured with higher accuracy than the other parameters, which should eventually be adjusted. Consistent orbital solutions and system parameter fits for the relevant systems, however, are beyond the scope of the present work.

The case of the planet OGLE2-TR-L9b provides an example of significantly inconsistent parameters. This planet is a relatively massive hot Jupiter orbiting a rapidly rotating F3 star in a circular orbit. The values of 1.52~$\pm$~0.08 $M_{\sun}$ and 0.0308~$\pm$~0.0005 AU for the mass of its host star and its orbital distance, respectively, in the exoplanet.eu catalog come from the discovery paper by \citet{snellen09}. These values give a Keplerian orbital period of 1.6~$\pm$~0.06 days that differs by 15.5$\sigma$ from the listed orbital period of 2.4855335~$\pm$~7~$\times$~10$^{-7}$ days. In contrast, the homogenous study of transiting systems by \citet{southworth10} gives values of 1.42~$\pm$~0.11 $M_{\sun}$ and 0.0404~$\pm$~0.0011 AU for the host star mass and orbital distance, respectively. These values give a Keplerian period of 2.485 days that agrees with the measured period. 

Another example of inconsistent parameters is the case of the planet NGTS-3Ab, a hot Jupiter orbiting the primary star of an unresolved binary in a circular orbit \citep{gunther18}. The listed values of 1.017~$\pm$~0.093 M$_{\sun}$ and 0.023$_{-0.0046}^{+0.0065}$ AU for the host star mass and orbital distance, respectively, give a Keplerian orbital period of 1.26~$\pm$~0.46 days whereas the measured orbital period is 1.68~$\pm$~3~$\times$~10$^{-6}$ days. In this case, the solution to the discrepancy is trivial because two values for the orbital period are given in the discovery paper by \citet{gunther18} and the incorrect value ended up in the catalog. The orbital distance is listed as $a =$~5$_{-1.0}^{+1.4}$ $R_{\sun}$ i.e., about 0.023 AU, but the actual orbital fit parameter is given as $R_p/a =$~0.02523~$\pm$~0.00071 that implies $a =$~0.028 AU. The latter value for the orbital distance gives a Keplerian period of 1.695 days that is consistent with the measured period.    

The parameters for the planets NGTS-4b and TOI-132b have relatively large uncertainties \citep{west19,diaz20}. These planets are interesting because they reside in the so-called hot Neptune desert, and the stability of their envelopes is of obvious interest to models of mass loss. The listed host star mass and orbital distance for NGTS-4b are 0.75~$\pm$~0.02 M$_{\sun}$, and 0.019~$\pm$~0.005 AU, respectively \citep{west19}. These values give a Keplerian orbital period of 1.1~$\pm$~0.44 days, which differs from the measured orbital period of 1.34~$\pm$~8~$\times$~10$^{-6}$ days. The uncertainty on the orbital distance, however, is relatively large and shifting the planet farther from the star to $a =$~0.0216 AU would give a Keplerian period that agrees with the measured orbital period. This value lies within 1$\sigma$ from the listed orbital distance and produces a stable solution for the shape of the planet based on the Roche potential. The listed host star mass and orbital distance for TOI-132b are 0.97~$\pm$~0.06 M$_{\sun}$, and 0.026$_{-0.003}^{+0.002}$ AU, respectively \citep{diaz20}. These values give a Keplerian period of 1.55~$\pm$~0.23 days, which is shorter than the listed orbital period of 2.1097008~$\pm$~0.00003 days by 2.4$\sigma$.  In this case, an orbital distance of 0.0318 AU would lead to an agreement between the Keplerian and measured orbital periods, and a stable shape for the planet.  

Finally, we verify that the hot Jupiters with the highest $X/Z$ ratios, WASP-12b, WASP-19b, and WASP-121b have consistent system parameters. For WASP-12b, the system parameters come from \citet{collins17}, with a slightly revised orbital semi-major axis from \citet{lanza20} that is consistent with the earlier value. The parameters are consistent with a Keplerian orbit to within 0.2\%. For WASP-121b, the system parameters are from \citet{delrez16} and they are consistent with a Keplerian orbit to within 0.1\%. For WASP-19b, the system parameters are from \citet{mancini13}. The Keplerian period is 0.82~$\pm$~0.02 days, which disagrees with the listed orbital period of 0.78884~$\pm$~0.0000003 by about 4\%. The orbits of these three planets are consistent with circular or at most, very small eccentricity. 

\section{Discussion}  
\label{sc:discussion}

\begin{figure}
  \epsscale{1.3}
  \plotone{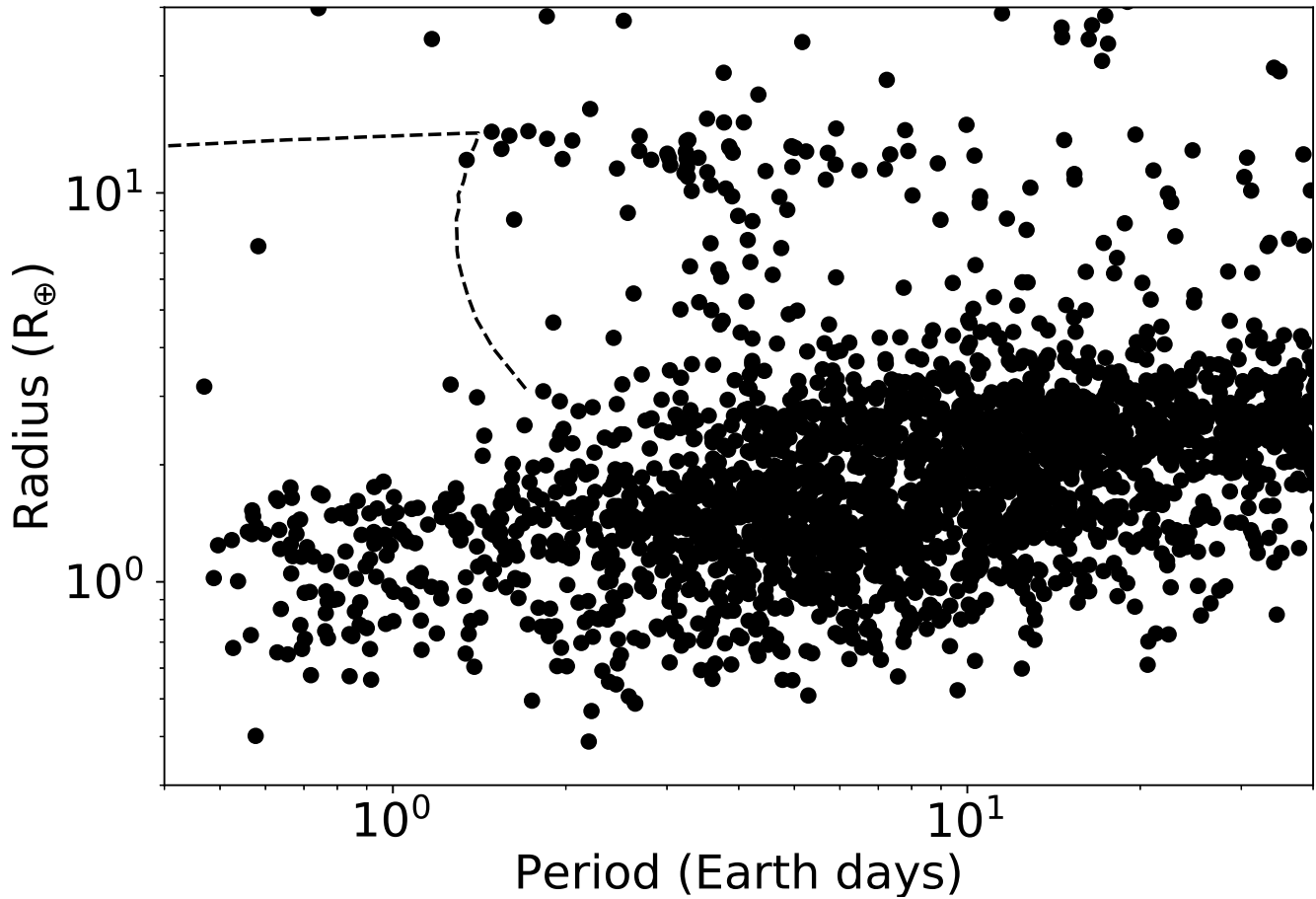}
  \caption{Planets orbiting main sequence stars detected by Kepler \citep[filled circles, based on revised parameters from][]{berger20} compared with the limit where the mass loss rate due to Roche lobe overflow exceeds the stellar XUV-driven escape rate for different planets orbiting a sun-like star (dashed line). Roche lobe overflow dominates at short orbital distances for planets with radii less than about 13--14 $R_{\Earth}$.}
  \label{fig:kep_pop}
\end{figure}

\noindent
As we note before, many studies of exoplanet evolution rely on energy-limited escape to represent photoevaporation. This approach underestimates mass loss rates due to Roche lobe overflow. Therefore, we explore the dependency of our results on an extended set of planet properties and relate them to the population of planets detected by Kepler. Figure~\ref{fig:kep_pop} shows the periods and radii of Kepler planets orbiting main sequence stars, following a revision of their properties based on stellar properties from the Gaia Data Release 2 and DR25 Kepler Stellar Properties Catalogue \citep{berger20}. The dashed line shows the limit where the minimum mass loss rate due to Roche lobe overflow exceeds energy-limited photoevaporation. In order to calculate this limit, we create a grid of planet masses from 10 $M_{\Earth}$ to 960 $M_{\Earth}$ with a spacing of 5 $M_{\Earth}$ and use the empirical mass-radius (M-R) relationship of \citet{chen17} to obtain radii for the planets. This relationship is mostly based on transiting planets that typically orbit close to their host stars. For a clear atmosphere, the visual transit probes the 0.01-0.1 bar level in the atmosphere, and we assume that the radius applies at the 0.1 bar level.

The Kepler population includes Ultra-short period, or USP, planets with periods of $P \le$~1 day. These planets generally have radii of less than about 2 $R_{\Earth}$ and they are unlikely to have significant gaseous envelopes. \citet{sanchisojeda14} suggested that larger USP planets lose their gaseous envelopes to photoevaporation. Our results confirm that larger USP planets can lose their envelopes to rapid Roche lobe overflow. The cutoff for catastrophic mass loss is at planet radii of 13--14 $R_{\Earth}$ that correspond to planet masses comparable to or slightly lower than the mass of Jupiter. The population of smaller USP planets can therefore include remnant cores of both sub-giant and giant planets. Curiously, there are two USP candidate planets with radii larger than 2 $R_{\Earth}$. These planets orbit stars with effective temperatures of 5744 K and 5889 K and radii of 1.34 and 1.25 $R_{\Sun}$, respectively. \citet{sanchisojeda14} argued that such planets may retain some of their gaseous envelopes if they have large enough core masses and densities. This would make them significant outliers in the M-R distribution, however, and thus the nature of these mysterious planets remains unclear. 

\citet{berger18} refine the limits of a hot Super-Earth and Neptune `desert', a dearth of planets with $R_p =$~2--4 $R_{\Earth}$ and host star bolometric fluxes of $F \gtrsim$~650 $F_{\Sun}$. The flux limit corresponds to an orbital distance of $a \approx$~0.04 AU ($P \approx$~3 days) for a planet orbiting a sun-like star. Based on the M-R relationship, the lowest mass planet in our grid has a radius of $R_p \approx$~3 $R_{\Earth}$. Extrapolating to slightly lower masses indicates that Roche lobe overflow can play a significant role in shaping the hot Neptune desert. This desert is a part of a larger sub-Jovian desert that extends to $P \approx$~4 days, including planets with radii of $R_p \approx$~4--10 $R_{\Earth}$ \citep[e.g.,][]{mazeh16}. Our results indicate that a part of this desert can also be cleared by Roche lobe overflow. We note that these results are likely to be conservative because gaseous planets have higher internal energies, larger radii, and thus lower bulk densities when they form, making them more sensitive to mass loss in their youth \citep[e.g.,][]{kurokawa14}. 

Beyond the limit for significant Roche lobe overflow, escape is powered mostly by the stellar XUV radiation and mass loss is roughly energy-limited as long as escape is hydrodynamic and the extent of the atmosphere is correctly represented. We cannot speculate on the consequences of escape on the planets in this regime as the results depend significantly on the assumed interior structure and envelope composition \citep[e.g.,][]{owen17}, the modeling of which is beyond the scope of this work. For the same reason, we have not explored the consequences of Roche lobe overflow with detailed models of planetary or orbital evolution in this work. 

Formally, energy-limited escape can be defined as mass loss that is directly proportional to the net energy deposited as heat in the upper atmosphere. We note that even when escape is energy-limited, the heating efficiency of the upper atmosphere remains a significant uncertainty in the models. Not only does the heating efficiency depend on the detailed composition of the upper atmosphere but it also depends on the incident stellar XUV flux \citep[e.g.,][]{murrayclay09}. As if this was not enough, the heating efficiency is also likely to depend on the spectral energy distribution of the XUV flux. For example, it is well known that young sun-like stars are much more active than the current sun and emit more of their XUV flux in X-rays that penetrate deeper to the lower thermosphere and middle atmosphere. Thus the heating and cooling profiles in the atmospheres of young planets can differ significantly from planets that orbit current sun-like stars and the heating efficiency is likely to change significantly over time.  

The key limitations of our models are that they are one-dimensional and only apply to hydrogen and helium envelopes. They also do not include the possibility of a significant planetary magnetic field or the interaction of the planetary upper atmosphere and exosphere with the stellar wind. The degree to which the differences in temperature and composition between the day and night sides and subsequent horizontal dynamics affect the mass loss rate is not immediately obvious. One-dimensional models are typically able to include more detailed physics and remain tractable while multi-dimensional models are often based on simplified radiative transfer and chemistry to provide a more realistic representation of the redistribution of energy and different species around the planet. This poses a key problem for objective comparisons between 1D and multi-dimensional models. It is often difficult to work out if the differences between models are due to horizontal dynamics or some other assumptions about chemistry, heating and cooling, lower atmosphere structure, or escape mechanism. In this context, recent work by \citet{guo13} and \citet{owen20} who compared different models with similar physics provide useful insights.

\citet{owen20} compared predictions by our model for the warm Neptune GJ436b \citep{loyd17} with similar but 3D simulations of \citet{shaikhislamov18}. There is a relatively good agreement on globally averaged atmospheric structure between the models and most importantly, the mass loss rates agree to within 10\%. We suspect that this is at least in part due to the fact that in extended upper atmospheres, stellar XUV radiation passes the terminator to the night side, leaving only a small region behind the planet in complete shadow \citep[e.g., see the lower right panel of Figure 2 in][]{koskinen07}. \citet{guo13}, on the other hand, compared results from 1D and 2D versions of his model to explore Roche lobe effects and temperature differences between the day and night side on a planet like HD209458b. In this case, the symmetry axis of the 2D model was aligned along the star-planet line from the substellar point to the anti-stellar point. The simulations show that a globally averaged 1D model with substellar gravity overestimates the mass loss rate by a factor of two compared to a 2D model with uniform heating around the planet. If one also includes the change in temperature between the day and night sides in the 2D model, the mass loss rate is about seven times smaller than the mass loss rate predicted by the globally averaged 1D model with substellar gravity. Note, however, that these differences could be overestimated because dynamics in the 2D model is still restricted and the difference between the 1D and 3D models compared by \citet{owen20} is smaller.

While many models with magnetic fields and stellar wind interaction have been developed for transiting planets \citep[e.g.,][]{ip04,erkaev05,lipatov05,preusse07,trammell11,owen14,strugarek14,matsakos15}, few if any models of exoplanet evolution include them. This is understandable, given that observational constraints on the nature and evolution of stellar winds and planetary magnetic fields outside of the solar system are weak or non-existent. Their effect on the planetary upper atmosphere and escape rates is also not fully understood. On one hand, a magnetic field may impede the escape of ions at low and middle latitudes and restrict it to polar winds \citep[e.g.,][]{trammell11}. If the ion-neutral collision frequency is sufficiently high, trapped ions may also impede the escape of the neutral atmosphere that is not otherwise impeded by the magnetic field. On the other hand, a planetary magnetic field, particularly when combined with the incident stellar wind, can lead to strong resistive (Joule) heating in the upper atmosphere \citep{cohen14}, not only in extended auroral zones but also in non-auroral regions where electric currents can be generated by a wind dynamo \citep{koskinen14}. This additional heating could substantially increase the escape rate, even if escape only takes place at middle and high latitudes.

\citet{koskinen13b} used arguments based on plasma $\beta$ (the ratio of the thermal pressure to magnetic pressure) and the second Cowling number Co (the ratio of magnetic pressure to the planetary wind dynamic pressure) to estimate the maximum magnetic field strength that would allow for the unimpeded escape of ions from a planet like HD209458b. They inferred a limiting magnetic moment of about 0.04 $m_J$, where $m_J =$~1.5~$\times$~10$^{27}$ A~m$^2$ is the magnetic moment of Jupiter, corresponding to an equatorial magnetic field strength of about 0.1 G at the 1 R$_p$ level. Similar limiting values for hot Jupiters have been inferred in subsequent work \citep[e.g.,][]{owen14}. The maximum allowed magnetic field strength, however, increases significantly when Roche lobe overflow sets in. For example, the thermal pressure at the L1 point in our Model B of a hot Uranus at 0.02 AU is 40 nbar while the dynamical pressure is 58 nbar. For the magnetic pressure to exceed these values, the required equatorial magnetic field strength at the surface pressure level is about 27 G (corresponding to about 1.2 G at the L1 point). This field is much stronger than the 1 bar equatorial magnetic field of 4.3 G on Jupiter, which is the strongest planetary magnetic field in the solar system.

The effects of different assumptions about composition and radiative transfer on the mass loss rate could be larger than the effect of multi-dimensional dynamics or magnetic fields that appear to be limited to a factor of a few or an order of magnitude at most. Generally, the extent of the atmosphere and the mass loss rate depend on the temperature and mean molecular weight. Our simplified approach reproduces the mean molecular weight of a solar composition atmosphere in chemical equilibrium. It ignores photochemistry in the middle atmosphere than can dissociate molecules more efficiently and produce a more extended atmosphere. The potential of photochemistry to significantly alter our results, however, is debatable, particularly for hotter atmospheres that are expected to be close to chemical equilibrium. The overall effect of the composition, particularly the presence of heavier atoms and molecules, on the heating and cooling rates, however, could alter the temperature profile in the middle and upper atmosphere. 

The effect of the composition is particularly important if the envelope metallicity is higher than solar. This is the case for the ice giants Uranus and Neptune in the solar system. Several studies also argue that super-solar metallicity is required to explain observations of transiting Super-Earths and warm Neptunes \citep[e.g.,][]{lavvas19}. Further support for this idea may be inferred from the empirical mass-radius relationship for exoplanets that is largely based on properties of transiting close-in planets \citep{chen17}. Interior structure models predict that a pure hydrogen and helium envelope of just a few percent of the planet mass around an Earth-like rock-metal core roughly doubles the radius of a Super-Earth \citep[e.g.,][]{owen17}. The mean slope of the observed mass-radius relationship above the rocky planet limit does not appear to be as steep as this would imply, suggesting that Super-Earths typically do not have pure hydrogen and helium or solar metallicity envelopes.

The potential of an envelope with significantly super-solar metallicity to lose mass to escape could be much lower than that of a pure hydrogen and helium envelope. This would not change the fundamental results or the theory in this work but it would lead to different quantitative outcomes for mass loss and stability limits. For example, from solar to 1000 times solar metallicity, the mean molecular weight of the atmosphere increases from 2.3 amu to 15--17 amu \citep{lavvas19}. This leads to smaller scale height and extent of the atmosphere. At the same time, temperatures in the middle and upper atmosphere could become cooler. Enhanced metallicity produces larger abundances of radiatively active molecules such as H$_2$O, CH$_4$, and CO$_2$ that can cool the atmosphere and thereby reduce mass loss rates. For example, the thermosphere of Venus is much cooler than the thermosphere of the Earth because it is primarily composed of CO$_2$, which is an efficient coolant unlike the homonuclear diatomic molecules N$_2$ and O$_2$ that dominate on the Earth.

\section{Summary and conclusions}

\noindent
In this work, we focus on the consequences of mass loss from the closest-in exoplanets, with an emphasis on the transition from stellar XUV-driven escape of the upper atmosphere to Roche lobe overflow of the middle atmosphere that is primarily heated by the stellar bolometric luminosity. This work was inspired by the NUV observations of WASP-12b and WASP-121b \citep{fossati10,sing19}, together with earlier work on related theory \citep[e.g.,][]{lecavelier04,li10}. The visible surface of a planet need not fill the Roche lobe for the planet to lose its envelope over a relatively short time \citep[see also][]{jackson17} and we quantify the limit for significant envelope loss in terms of atmospheric structure and exoplanet system properties. It occurs once the 10$^{-8}$--10$^{-7}$ bar level in the upper atmosphere reaches the Roche lobe. When this happens, stellar XUV heating and photoevaporation driven by it are no longer important because the optical depth to ionizing radiation in the XUV reaches unity at around the 10$^{-9}$ bar level.

The equilibrium shape based on the Roche potential can be used to predict the stability of gaseous envelopes. We represent the equilibrium shape of close-in planets by calculating the ratio of the substellar radius to the polar radius ($X/Z$) at the level in the atmosphere that would be probed by visual broadband transit observations. At the Roche lobe, this ratio would be about 1.5. For a Jupiter-like planet orbiting a sun-like star, a ratio of $X/Z \sim$~1.2 implies rapid envelope loss. The same limit for a Uranus-like planet orbiting a sun-like star is only $X/Z \sim$~1.02. This is because the Uranus-like planet has a lower surface gravity and relatively more extended envelope than hot Jupiters. The latter conclusion also applies more generally to Mini-Neptunes and Super-Earths. These results are supported by the properties of known exoplanets. Among the known transiting planets for which we have roughly reliable masses and radii, there are none with implied $X/Z$ ratios that exceed 1.2. In addition, all planets with masses less than 100 $M_{\Earth}$ have $X/Z$ ratios close to unity while the few planets with higher ratios are all hot Jupiters.

As expected, the planets WASP-12b and WASP-121b have some of the highest $X/Z$ ratios in our sample. This explains the detection of substantial mass loss that includes metals and heavy elements from these planets \citep{fossati10,sing19}. In fact, the surface distortion of these planets is so significant that Roche lobe effects should be taken into account in studies of their lower and middle atmospheres and are required for the interpretation of the entire UV--IR transmission spectrum. How fast these planets might lose their envelopes, however, is subject to debate. This is because the mass loss rate can vary substantially with atmospheric structure and model assumptions for planets that approach the stability limit. What is clear is that the mass loss rate from these planets exceeds the stellar XUV-driven escape rate and that their upper atmospheres are fundamentally different from more `mainstream' hot Jupiters.   

Many studies of exoplanet evolution rely on simple energy-limited escape to represent stellar XUV-driven atmospheric escape \citep[e.g.,][]{owen17}. This approach significantly underestimates Roche lobe effects that cannot fully be captured by adding a simple correction to energy-limited escape based on the potential difference between the surface pressure level of the planet and the L1 point \citep{erkaev07}. Studies of exoplanet evolution using this approach underestimate the effect of Roche lobe overflow that is likely to play a significant role in clearing out the Super-Earth and hot Neptune deserts at the shortest orbits. Our results indicate that Roche lobe overflow can effectively clear out the sub-Jovian desert at orbital periods of less than about two days, even if we do not account for the early evolution of the planets when mass loss rates are likely to be even higher. We look forward to future studies of exoplanet populations that incorporate these results in more complete, coupled models of mass loss and evolution. 

We use a simple approach to represent the extent of the lower and middle atmosphere and calculate the mass loss rate due to Roche lobe overflow under the assumption that the atmosphere escapes through the L1 point roughly at the speed of sound. Our detailed simulations of atmospheric escape with a one-dimensional multi-species model support this approach. The simulations also indicate that stellar XUV-driven hydrodynamic escape is consistent with energy-limited escape as long as the models properly account for the extent of the upper atmosphere where XUV radiation is absorbed and escape is initiated. By energy-limited escape here we mean mass loss rates that are directly proportional to the net energy deposited as heat in the upper atmosphere. The heating efficiency of the upper atmosphere persists as a key uncertainty in this regime. 

The main limitations of our model are the fact that they are one-dimensional, only apply to hydrogen and helium envelopes, and do not include a possible planetary magnetic field or interaction of the planetary upper atmosphere with the stellar wind. Comparisons of 1D, 2D, and 3D models with similar physics indicate that the differences in the global mass loss rates predicted by models with different dimensions are at most a factor of a few \citep{guo13,owen20} while the effect of magnetic fields and stellar winds may also be of secondary importance to thermal escape for the closest-in exoplanets. Possible differences in envelope composition may be a more significant source of uncertainty in escape models than multi-dimensional dynamics at this point. Ice giants in the solar system have super-solar envelope metallicities and observations of warm Neptunes and Super-Earths point to the same direction for lower-mass exoplanets. Super-solar metallicity leads to higher mean molecular weight in the atmosphere and likely more efficient radiative cooling, both of which reduce mass loss rates. The solar system also provides another warning for modelers of exoplanet escape: the atmosphere of Pluto.

Most models of atmospheric escape that were adapted to Pluto prior to the New Horizons (NH) flyby, including the one used in this study, predicted energy-limited escape with a heating efficiency of about 30\% from Pluto's extended N$_2$ and CH$_4$ atmosphere \citep{koskinen15}. Instead, the flyby observations revealed that the outer layers of the atmosphere are much cooler than expected and the escape rate is two orders of magnitude lower and consistent with the Jeans escape regime \citep{young18}. The reason for the lower than expected temperature is still not well understood but it could be due to an unforeseen radiative coolant, such as supersaturated water or aerosols \citep{strobel17,zhang17}. While close-in exoplanets obviously need not bear any resemblance to Pluto, we ought to allow for the possibility of surprises such as this, especially if the envelope metallicity is super-solar. 

In the end, future observations of exoplanet atmospheres will help to elucidate the composition of planetary envelopes and therefore their potential for mass loss over time. The extent of the hydrogen and helium envelope, as observed in the H~Lyman~$\alpha$ and He I lines, appears to be relatively larger for warm Neptunes than for hot Jupiters \citep[e.g.,][]{ehrenreich15,bourrier18}. Our model of a Uranus-like planet orbiting a sun-like star agrees with this general idea, showing that the relative extent of the cloud of neutral hydrogen surrounding the planet is much larger than on hot Jupiters. This is because the atmosphere is more extended and outflow has a more pronounced effect on the density structure on planets with a lower gravitational potential. In contrast to hot Jupiters, we also predict that H$_2$ molecules should be present at large distances from the planet, along with the molecular ions H$_3^+$ and HeH$^+$. While the detection of the molecular ions may prove difficult \citep[e.g.,][]{chadney16}, the detection of H$_2$ in the FUV could provide an interesting additional test for future observations \citep{barthelemy07} to determine if a hydrogen and helium envelope accurately represents the envelopes of hot Neptunes, Mini-Neptunes, and Super-Earths.

\begin{acknowledgments}
T. T. K. and C. H. acknowledge support by the NASA Exoplanet Research Program grant 80NSSC18K0569. R. B. F. and G. B. acknowledge support by NASA under grant 80NSSC21K0593 for the program `Alien Earths'. This work benefited from collaborations and/or information exchange within NASA's Nexus for Exoplanet System Science (NExSS) research coordination network sponsored by NASA's Science Mission Directorate. M.E.Y. acknowledges funding from the European Research Council (ERC) under the European Union's Horizon 2020 research and innovation program under grant agreement No 805445. We thank Ilaria Pascucci for helpful discussions on planet populations and evolution.
\end{acknowledgments}


%\vspace{5mm}
%\facilities{HST(STIS), Swift(XRT and UVOT), AAVSO, CTIO:1.3m,


%\software{astropy \citep{2013A&A...558A..33A,2018AJ....156..123A},  
%          Cloudy \citep{2013RMxAA..49..137F}, 
%          Source Extractor \citep{1996A&AS..117..393B}
%          }



\appendix

\section{Equipotential surfaces}
\label{ap:equipotentials}

\noindent
We calculate equipotential surfaces based on the Roche potential numerically by using the iterative method adapted from \citet{lindal85}. First, we set the reference potential at the north pole to:
\begin{equation}
U_p = U(z,0,0).
\end{equation}
Then we calculate the radii along the equipotential surface at different latitudes and longitudes by using an iterative method. For each point, we choose an initial guess $r_1 (\theta, \phi)$. The subsequent error in the potential at a given point is
\begin{equation}
\Delta U_1 =  U(r_1,\theta,\phi)-U_p.
\end{equation}
Based on this, the updated value $r_2$ is
\begin{equation}
r_2 = r_1 + \Delta r_1
\end{equation}
where
\begin{equation}
\Delta r_1 = \frac{\Delta U_1}{g_r}
\end{equation}
and
\begin{equation}
g_r = -\frac{\partial U}{\partial r}.
\end{equation}
The iteration concludes once $\Delta U$ is vanishingly small, usually after just a few steps.

\section{CETIMB: An upper atmosphere escape model}
\label{ap:atmosmodel}

\noindent
CETIMB is a one-dimensional model of the thermosphere-ionosphere and hydrodynamic escape for close-in exoplanets. The model was described and applied to the hot Jupiter HD209458b by \citet{koskinen13a,koskinen13b}. Here, we provide a revised description that includes recent updates and features specific to the current study. The model solves the coupled equations of continuity, momentum, and energy in the radial direction:
\begin{eqnarray}
\frac{\partial \rho_s}{\partial t} + \frac{1}{r^2} \frac{\partial}{\partial r} \left[ r^2 \rho_s (w + w_s) \right] &=& \sum_t \rho_s R_{st} \\
\frac{\partial (\rho w)}{\partial t} + \frac{1}{r^2} \frac{\partial}{\partial r} \left( r^2 \rho w^2 + p \right) &=& -\rho \frac{\partial U}{\partial r} + F_{\mu} \\
\frac{\partial (\rho u)}{\partial t} + \frac{1}{r^2} \frac{\partial}{\partial r} \left( r^2 \rho u w \right) &=& \rho q - p \frac{1}{r^2} \frac{\partial}{\partial r} \left( r^2 w \right) + \frac{1}{r^2} \frac{\partial}{\partial r} \left( r^2 \kappa \frac{\partial T}{\partial r} \right) + q_{\mu} \label{eqn:ener1} \\
\rho = \sum_s \rho_s, \ \ w &=& \frac{1}{\rho} \sum_s \rho_s (w + w_s) \\
\end{eqnarray}
where $\rho_s$ is the mass density of species $s$, $w+w_s$ is the species velocity, $w$ is the center of mass velocity, $R_{st}$ is the chemical reaction rate for species $s$ with species $t$, $p$ is pressure, $U$ is the gravitational potential, $u = c_v T$ is the specific internal energy, $q$ is the sum of radiative heating and cooling rates, and $\kappa$ is the coefficient of heat conduction. The viscous momentum flux and dissipation functional, respectively, are:
\begin{eqnarray}
F_{\mu} &=& \frac{4}{3} \frac{1}{r^2} \frac{\partial}{\partial r} \left( r^2 \mu \frac{\partial w}{\partial r} \right) - \frac{\partial \mu}{\partial r} \frac{\partial w}{\partial r} - \frac{16}{3} \mu \frac{w}{r^2} \\
q_{\mu} &=& \frac{10}{3} \mu \left( \frac{\partial w}{\partial r} \right)^2 - \frac{8}{3} \mu \frac{w}{r} \frac{\partial w}{\partial r} + \frac{16}{3} \mu \left( \frac{w}{r} \right)^2
\end{eqnarray}
where $\mu$ is the coefficient of viscosity. In this study, the simulations use either the normal, spherical gravitational potential (Model~A) or the Roche potential for the substellar streamline (Model B):
\begin{equation}
\phi = -\frac{GM_p}{r} - \frac{GM_s}{a-r} - \frac{1}{2} \Omega^2 \left(r - M_{sp} a \right)^2
\end{equation} 
where $M_p$ is the mass of the planet, $M_s$ is the mass of the host star, $a$ is the orbital distance, 
\begin{equation}
M_{sp} = \frac{M_s}{M_s+M_p},
\end{equation}
and $\Omega$ is the orbital angular frequency. 

As indicated above, the model solves separate continuity equations for the different neutral and ion species coupled to common momentum and energy equations for the bulk atmosphere. The use of the diffusion approximation is therefore necessary to solve for the individual species velocity perturbation $w_s$. The diffusion velocities are obtained by matrix inversion from
\begin{equation}
\left( 1 + \Lambda_s \right) \frac{\partial x_s}{\partial r} + \left( x_s - \frac{\rho_s}{\rho} \right) \frac{\partial (\ln p)}{\partial r} - \frac{n_s e_s E}{p} = -\sum_{t \ne s} \frac{x_s x_t}{D_{st}} \left( w_s - w_t \right) 
\end{equation}
under the condition that
\begin{equation}
\sum_s \rho_s w_s = 0,
\end{equation}
where $x_s$ is the volume mixing ratios, $\Lambda_s = K_{zz}/D_s$, $K_{zz}$ is the eddy diffusion coefficient, 
\begin{equation}
\frac{1}{D_s} = \sum_{t \ne s} \frac{x_t}{D_{st}}
\end{equation}
is the net molecular diffusion coefficient, $D_{st}$ is the mutual diffusion coefficient, $e_s$ is the electric charge, and $E$ is the radial electric field. We assume that the planetary magnetic field is negligible and ignore thermal diffusion since the coefficients for the latter are poorly known at the relevant temperatures. The diffusion coefficients, that are based on momentum transfer collision frequencies, include neutral-neutral, resonant and non-resonant ion-neutral, and electron-ion interactions. The diffusion approximation does not place constraints on the bulk (center of mass) velocity $w$ that can be either subsonic or supersonic. It does place constraints on the magnitude of $w_s$ but it can be shown that a violation of the conditions in which the approximation is valid is very unlikely to occur and generally requires minor species velocities to be significantly faster than the bulk flow velocity. 

In this work, we include the species H$_2$, H, He, H$_2^+$, H$_3^+$, H$^+$, He$^+$, and HeH$^+$. While it is possible and perhaps likely that the envelopes of hot Neptunes, Mini-Neptunes and Super-Earths have enhanced metallicities, with molecules and heavy atoms extending to the upper atmosphere, we do not include elements heavier than H and He. Our calculations here are primarily provided to test the common approximations and simplifications in Sections~\ref{subsc:overflow1} and \ref{subsc:energy_limit}. The relevant chemical reactions used by the model are given in Table~\ref{table:reactions} with references. The ionization rates are calculated self-consistently by the model based on the mean solar spectrum and relevant cross sections. We note that our model does not include neutral dissociation of H$_2$ by photons with wavelengths longer than the ionization threshold of 80.35 nm (15.43 eV). The cross section of H$_2$ at these longer wavelengths consists of a multitude of electronic bands and has to be computed at high resolution \citep[e.g.,][]{koskinen21}. Excitation of the electronic bands leads to the emission of fluorescent radiation and dissociation, where the probability for dissociation is 0.1-0.15 \citep{shematovich10}. 

\begin{deluxetable*}{lcc}
  \tablecolumns{3}
  \tablecaption{Reaction rate coefficients} 
  \tablehead{
    \colhead{Reaction}  & 
    \colhead{Rate (cm$^3$ s$^{-1}$)}  & 
    \colhead{Reference} 
    }
  \startdata
  P1 \ \ \ \  H$\ + \ h\nu \ \rightarrow$~H$^+ \  +   e$                                & SC & \citet{hummer63} \\
  P2 \ \ \ \  He$\ + \ h\nu \ \rightarrow$~He$^+ \  +   e$                            & SC & \citet{yan98} \\
  P3 \ \ \ \  H$_2\ + \ h\nu \ \rightarrow$~H$_2^+ \  +   e$                        & SC & \citet{yan98} \\
  P4 \ \ \ \  H$_2\ + \ h\nu \ \rightarrow$~H$^+ \  +$ H $+\  e$                 & SC & \citet{chung93} \\
  P5 \ \ \ \  H$_2\ + \ h\nu \ \rightarrow$~H$^+ \  +$ H$^+ +  e + e$        & SC & \citet{dujardin87} \\
  R1 \ \ \ \  H$^+ + e \rightarrow$~H~$ + \ h\nu$                                     & 4.0~$\times$~10$^{-12} (300/T_e)^{0.64}$                                                                                      & \citet{storey95}  \\
  R2 \ \ \ \  He$^+ + e \rightarrow$~He~$ + \ h\nu$                                 & 4.6~$\times$~10$^{-12} (300/T_e)^{0.64}$                                                                                      & \citet{storey95}  \\
  R3 \ \ \ \  H$\  +\  e \rightarrow$~H$^+ + e + e$                                    & 2.91~$\times$~10$^{-8} U^{0.39} \exp(-U)/(0.232+U) \ , \ U =  13.6/E_e(eV)$                                & \citet{voronov97}  \\
  R4 \ \ \ \  He$\  +\  e \rightarrow$~He$^+ + e + e$                                & 1.75~$\times$~10$^{-8} U^{0.35} \exp(-U)/(0.180+U) \ , \ U =  24.6/E_e(eV)$                                & \citet{voronov97} \\ 
  R5 \ \ \ \  H$_2^+ + e \rightarrow$~H + H                                              & 2.3~$\times$~10$^{-8} (300/T_e)^{0.4}$                                                                                           & \citet{auerbach77} \\   
  R6 \ \ \ \  H$_3^+ + e \rightarrow$~H$_2$ + H                                      & 2.16~$\times$~10$^{-8} (300/T_e)^{0.65}$                                                                                       & \citet{larsson08} \\ 
  R7 \ \ \ \  H$_3^+ + e \rightarrow$~H + H + H                                       & 5.04~$\times$~10$^{-8} (300/T_e)^{0.65}$                                                                                       & \citet{larsson08} \\   
  R8 \ \ \ \  H$_2^+ +$ H$_2 \rightarrow$~H$_3^+$ + H                         & 2~$\times$~10$^{-9}$                                                                                                                       & \citet{thread74} \\   
  R9 \ \ \ \  H$_2^+ +$ H$ \rightarrow$~H$^+$ + H$_2$                         & 6.4~$\times$~10$^{-10}$                                                                                                                  & \citet{kapras79} \\       
  R10 \ \ \ \  H$^+ +$ H$_2 (\nu \ge 4) \rightarrow$~H$^+$ + H$_2$     & 10$^{-9}$ $\exp(-21900/T)$                                                                                                               & \citet{yelle04} \\  
  R11 \ \ \ \  H$_3^+ +$ H$ \rightarrow$~H$_2^+$ + H$_2$                   & 2.1~$\times$~10$^{-9}$ $\exp(-20000/T)$                                                                                        & \citet{harada10} \\ 
  R12 \ \ \ \  H$_2 +$ M$ \rightarrow$~H + H + M                                   & 1.5~$\times$~10$^{-9}$ $\exp(-48350/T)$                                                                                        & \citet{baulch92} \\ 
  R13 \ \ \ \  H$^+ +$ H$_2$ + M$ \rightarrow$~H$_3^+$ + M               & 3.2~$\times$~10$^{-29} n$                                                                                                                & \citet{miller68} \\
  R14 \ \ \ \  H$_2 +$ e$ \rightarrow$~H + H + e                                     & 1.33~$\times$~10$^{-6} (300/T_e)^{0.91}$ $\exp(-55800/T)$                                                          & \citet{stibbe99} \\   
  R15 \ \ \ \  H + H + M$ \rightarrow$~H$_2$ + M                                   & 8~$\times$~10$^{-33} (300/T_e)^{0.6} n$                                                                                         & \citet{ham70} \\  
  R16 \ \ \ \  HeH$^+$ + e$ \rightarrow$~He + H                                     & 10$^{-8} (300/T_e)^{0.6}$                                                                                                                 & \citet{yousif89} \\ 
  R17 \ \ \ \  He$^+ +$ H$_2 \rightarrow$~H$^+$ + H + He                     & 10$^{-9}$ $\exp(-5700/T)$                                                                                                                & \citet{moses00} \\  
  R18 \ \ \ \  HeH$^+ +$ H$_2 \rightarrow$~H$_3^+$ + He                     & 1.5~$\times$~10$^{-9}$                                                                                                                    & \citet{bohme80} \\  
  R19 \ \ \ \  HeH$^+ +$ H$ \rightarrow$~H$_2^+$ + He                         & 9.1~$\times$~10$^{-10}$                                                                                                                  & \citet{kapras79} \\ 
  R20 \ \ \ \  He$^+ +$ H$_2 \rightarrow$~HeH$^+$ + H                         & 4.2~$\times$~10$^{-13}$                                                                                                                  & \citet{schauer89} \\                     
  R21 \ \ \ \  H$ \ +$~He$^+ \rightarrow$~H$^+ \  + \ $He                       & 1.2~$\times$~10$^{-15} (300/T)^{-0.25}$                                                                                          & \citet{stancil98}  \\
  R22 \ \ \ \  H$^+ \ +$~He$\  \rightarrow$~H$ \  + \ $He$^+$                 & 1.75~$\times$~10$^{-11} (300/T)^{0.75} \exp(-128,000/T)$                                                              & \citet{glover07}  \\
  R23 \ \ \ \  H$_2$ + He$^+ \rightarrow$~H$_2^+$ + He                       & 7.2~$\times$~10$^{-15}$                                                                                                                   & \citet{barlow84}
   \enddata
  \tablenotetext{}{SC: Calculated self-consistently based on ionization cross sections, stellar flux, and density profiles in the model.} 
  \label{table:reactions}
\end{deluxetable*}  

\begin{figure}
  \epsscale{0.75}
  \plotone{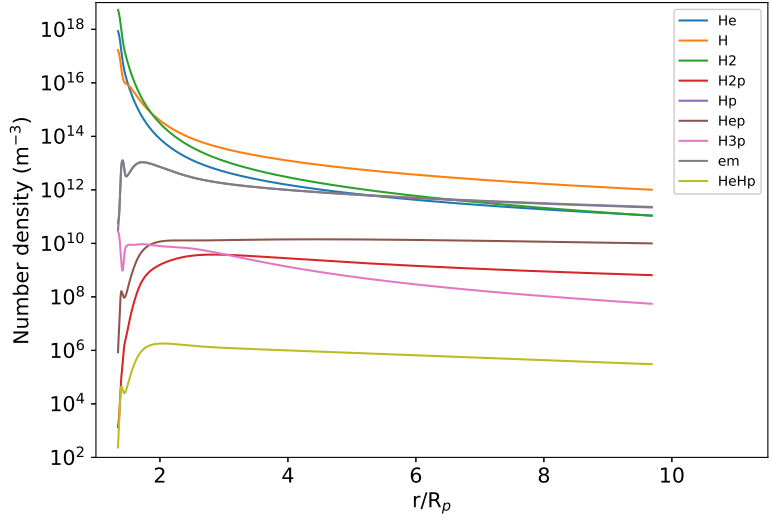}
  \caption{The composition for revised Model A at 0.05 AU that includes an approximate treatment of H$_2$ neutral dissociation and related heating (see text). The results are qualitatively similar to the model that does not include neutral dissociation of H$_2$ (see Figure~\ref{fig:compoA}).}
  \label{fig:apcompo}
\end{figure}

In our models of hot Jupiters, we have generally assumed that thermal dissociation of H$_2$ (reaction R12) and dissociation of H$_2$ due to photochemistry in the middle atmosphere \citep{moses11,lavvas14} are more important than dissociation of the electronic excited states. The same is not necessarily true for Neptunes with cooler atmospheres. The extension of the model to properly include the high resolution H$_2$ cross section and detailed calculations of neutral dissociation and the related heating rate requires more research and is beyond the scope of this work. However, we used a cross section based on the laboratory work of \citet{backx76} with a dissociation probability of 0.125 in the model to check the potential of neutral dissociation to change our results. Since this cross section has very low wavelength resolution, it is likely to overestimate the dissociation rate \citep[e.g., see Figure 1 in][]{kim14}. Figure~\ref{fig:apcompo} shows the resulting composition for Model A at 0.05 AU that should be compared with Figure~\ref{fig:compoA} for the model that does not include neutral dissociation of H$_2$. The abundance of H$_2$ is somewhat reduced in the revised model and the mass loss rate is slightly higher but the results are qualitatively similar and the mass loss rates agree to within a factor of 1.4. We have therefore decided to postpone the inclusion of detailed H$_2$ calculations to future work.   

\begin{figure}
  \epsscale{0.75}
  \plotone{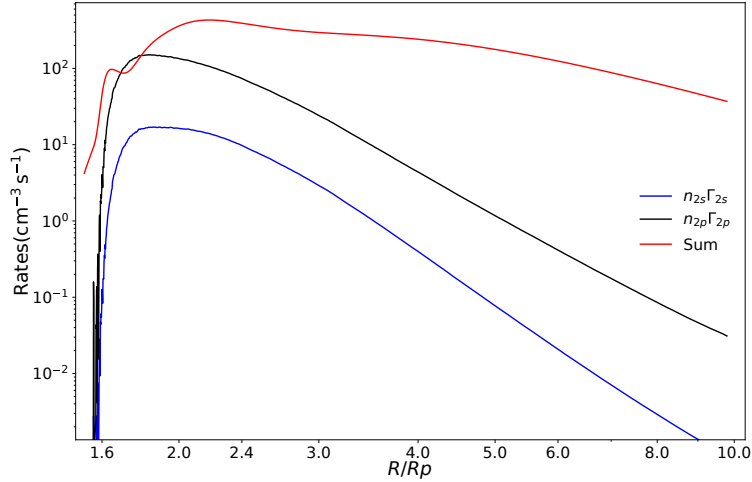}
  \caption{The H$^+$ production rates from photoionization of the H $n=2$ states ($n_{2s} \Gamma_{2s}$ and $n_{2p} \Gamma_{2p}$) compared with the sum of all included H$^+$ production rates (see Table~\ref{table:reactions}) based on Model B at 0.03 AU.}
  \label{fig:apbalmer}
\end{figure}

We also do not include direct ionization of H from the $n = 2$ state. This ionization pathway and related heating by the stellar flux in the Balmer continuum are potentially important for hot Jupiters \citep{huang17,garciamunoz19}. In order to test the possible significance of the $n = 2$ state to our results, we used the detailed balance model of \citet{huang17} to calculate the $2p$ and $2s$ excited state populations and the related ionization and heating rates for the simulations presented in Section~\ref{subsc:num_model}. In all cases, the excitation of H is dominated by the diffuse Lyman~$\alpha$ radiation field that arises from multiple resonant scattering of the stellar Lyman~$\alpha$ flux in the atmosphere. This poses a complex problem that we solve by using a Monte Carlo approach. We found the heating rate by the Balmer continuum to be negligible. The ionization rate, however, becomes comparable to and exceeds the ionization rate from the ground state by XUV radiation in a narrow region near the bottom of the model at 0.03 AU (see Figure~\ref{fig:apbalmer}). The impact of this additional ionization on our models, however, is small, and in reality, heavier molecules that might dominate the chemistry instead are likely to be present in this region. We have therefore also postponed the coupling of the escape models to the detailed balance model for hydrogen level populations to future work.  

\begin{figure}
  \epsscale{0.75}
  \plotone{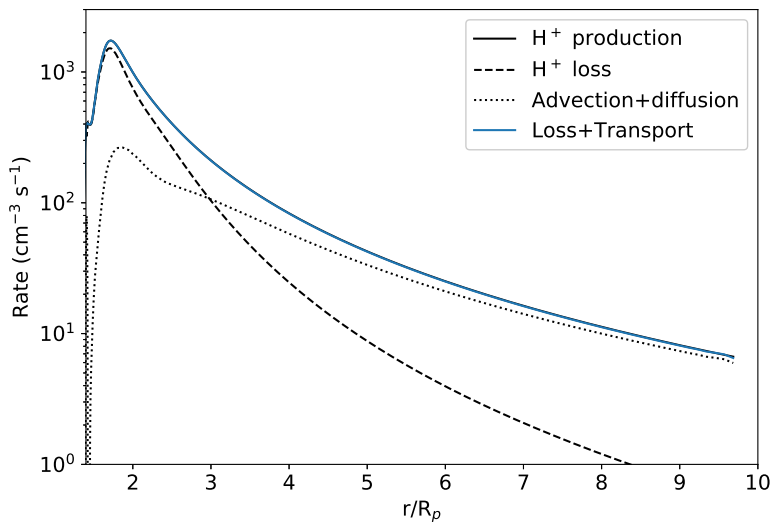}
  \caption{Production, loss, and transport terms (absolute values) in the continuity equation for protons in Model A at 0.05 AU.}
  \label{fig:aphcont}
\end{figure}

The prevalence of neutral H and outflow of H$_2$ at high altitudes around hot Neptunes differ from similar models of hot Jupiters. This leads to large transit depths at H~Lyman~$\alpha$ \citep{ehrenreich15,parkeloyd17} and may enable the future detection of H$_2$ on exoplanets \citep[e.g.,][]{barthelemy07}. To investigate the mechanism that enables these features, we analyze the terms of the continuity equation in detail for Model A at 0.05 AU. For example, Figure~\ref{fig:aphcont} shows the continuity equation terms for neutral H. At the lowest altitudes in the model, the chemical loss and production rates (ionization and recombination reactions) are in rough balance but above the heating peak, the ionization rate exceeds the recombination rate. Instead of recombination, ionization of H is balanced by upward transport of neutral H from below by escaping gas. In other words, outflow replenishes the outer layers of the atmosphere with neutral H from below, and keeps the upper atmosphere mostly neutral. The remarkable numerical accuracy of the model is highlighted by the near perfect alignment of the proton production rate (solid black line) with the sum of the proton loss and transport rates (solid blue line).    

\begin{figure}
  \epsscale{0.75}
  \plotone{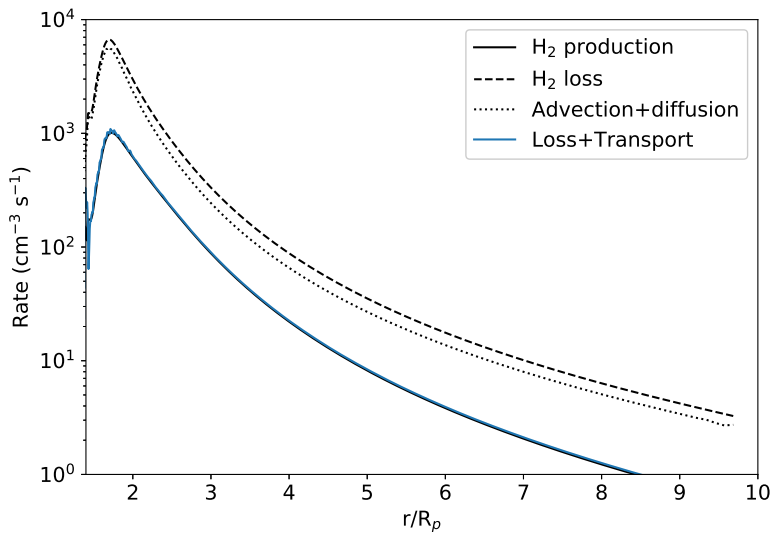}
  \caption{Production, loss, and transport terms (absolute values) in the continuity equation for H$_2$ in Model A at 0.05 AU.}
  \label{fig:aph2cont}
\end{figure}

We find a similar result for H$_2$, as illustrated by Figure~\ref{fig:aph2cont} that shows the continuity equation terms for H$_2$. Chemical loss of H$_2$ (dissociation and dissociative ionization reactions) dominates over production but outflow brings fresh H$_2$ up from deeper layers of the atmosphere and primarily balances the total loss rate. In the end, the sum of the chemical loss and transport terms equals the production rate, as expected in a steady state model. These remarkable results highlight the fact that escape is not a trivial process that can be ignored or simply parameterized in models that attempt to fit, say, exoplanet transmission spectra. Instead, it can fundamentally alter the ionization balance and relative abundances of different species in the upper atmosphere.

The sum of the radiative heating and cooling rates ($q$ in equation~\ref{eqn:ener1}) includes heating due to photoionization of the atmosphere by stellar XUV radiation and cooling by recombination, infrared emission by H$_3^+$, and H~Lyman~$\alpha$ emission. We calculate stellar XUV energy deposition self-consistently based on the density profiles in the model every 10 time steps, where one time step is one second. Each ionization event by a photon with frequency $\nu$ releases the energy $E = h \nu - I$ where $I$ is the ionization potential. In line with many previous studies \citep[e.g.,][]{murrayclay09}, we assume that 100\% of this energy heats the atmosphere and that the energy $I$ is lost in subsequent radiative recombination events or chemistry. In addition, we account for the loss of electron thermal energy in recombination of H$^+$ (the corresponding cooling rate due to the recombination of other ions is small). We use the H~Lyman~$\alpha$ cooling rate from \citet{glover07}, which is similar to the widely adapted rate coefficient from \citet{murrayclay09}, with the exception of a small temperature-dependent correction factor. We assume optically thin infrared cooling to space by H$_3^+$. The per molecule cooling rate is based on line lists \citep{neale96,miller13} and a correction factor for non-LTE conditions that is derived from detailed balance calculations \citep{koskinen09}. 

The model extends from the bottom of the thermosphere to several planetary radii. There are 410 radius levels on a stretched grid that mimics the expected spreading of pressure levels with altitude:
\begin{equation}
r_n = r_1 + \frac{\delta r_1 (1-s^n)}{1-s}
\end{equation}
where $n$ is the grid index of level $n$, $r$ is radius, $r_1$ is the radius at the lower boundary, $\delta r_1 =$~10 km is the grid spacing at the bottom, and $s =$~1.014 is the stretch factor. The lower boundary conditions are temperature, pressure ($p_1 =$~1 $\mu$bar), and species mixing ratios that are based on our lower atmosphere models. The upper boundary conditions are adaptable. If an exobase is found within the grid, the model uses modified Jeans escape conditions for individual species velocities, densities, and energy flux \citep{koskinen15} that incorporate the ambipolar electric field for ions \citep{koskinen13a} and Roche potential. If the exobase is above the sonic point, L1 point, or the grid, the model uses extrapolated boundary conditions appropriate for hydrodynamic escape \citep[e.g.,][]{tian05}. The model checks for the presence of an exobase every time step based on the collision frequencies between different species. All simulations in this paper are consistent with hydrodynamic escape.

\bibliography{koskinenms}{}
\bibliographystyle{aasjournal}



\end{document}